\documentclass[twocolumn,showpacs,preprintnumbers,amsmath,amssymb]{revtex4}
\usepackage{dcolumn}
\usepackage{color}
\usepackage{graphicx}

  \font\elevenmib=cmmib10 scaled 1095
 \font\tenmib=cmmib10
  \font\eightmib=cmmib10 scaled 800
  \font\sixmib=cmmib10 scaled 667
  \skewchar\elevenmib='177
  \newfam\mibfam
  \def\mib{\fam\mibfam\tenmib}
  \textfont\mibfam=\tenmib
  \scriptfont\mibfam=\eightmib
  \scriptscriptfont\mibfam=\sixmib
  \mathchardef\alpha="710B
  \mathchardef\beta="710C
  \mathchardef\gamma="710D
  \mathchardef\delta="710E
  \mathchardef\epsilon="710F
  \mathchardef\zeta="7110
  \mathchardef\eta="7111
  \mathchardef\theta="7112
  \mathchardef\kappa="7114
  \mathchardef\lambda="7115
  \mathchardef\mu="7116
  \mathchardef\nu="7117
  \mathchardef\xi="7118
  \mathchardef\pi="7119
  \mathchardef\rho="711A
  \mathchardef\sigma="711B
  \mathchardef\tau="711C
  \mathchardef\phi="711E
  \mathchardef\chi="711F
  \mathchardef\psi="7120
  \mathchardef\omega="7121
  \mathchardef\varepsilon="7122
  \mathchardef\vartheta="7123
  \mathchardef\varrho="7125
  \mathchardef\varphi="7127
    \def\physgreek{
    \mathchardef\Gamma="7100
    \mathchardef\Delta="7101
    \mathchardef\Theta="7102
    \mathchardef\Lambda="7103
    \mathchardef\Xi="7104
    \mathchardef\Pi="7105
    \mathchardef\Sigma="7106
    \mathchardef\Upsilon="7107
    \mathchardef\Phi="7108
    \mathchardef\Psi="7109
    \mathchardef\Omega="710A}

\def\etc{{\it etc.\/}}
\def\etal{{\it et al.\/}}

\def\ie{{\it i.e.\/}}

\def\sss#1{{\scriptscriptstyle #1}}

\def\ssr#1{{\sss{\rm #1}}}

\def\ONE{{1\kern-0.60em 1}}
\def\dsl{\raise.15ex\hbox{$/$}\kern-.57em\hbox{$\partial$}}
\def\nsl{\raise.15ex\hbox{$/$}\kern-.57em\hbox{$\nabla$}}
\def\id{\raise.72ex\hbox{$-$}\kern-.85em\hbox{$d$}\,}
\def\gtwid{\,{\raise.3ex\hbox{$>$\kern-.75em\lower1ex\hbox{$\sim$}}}\,}
\def\ltwid{\,{\raise.3ex\hbox{$<$\kern-.75em\lower1ex\hbox{$\sim$}}}\,}
\def\undr{\raise.3ex\hbox{$\sim$\kern-.75em\lower1ex\hbox{$|\vec x|\to\infty$}}}

\def\frac#1#2{{\textstyle{#1 \over #2}}}
\def\half{\frac{1}{2}}

\def\({\left (}
\def\){\right )}

\def\uar{\uparrow}
\def\dar{\downarrow}

\def\cE{{\cal E}}

\def\cI{{\cal I}}

\def\cK{{\cal K}}

\def\cM{{\cal M}}

\def\cO{{\cal O}}

\def\cQ{{\cal Q}}

\def\cS{{\cal S}}

\def\bfA{{\mib A}}
\def\bfB{{\mib B}}

\def\bfK{{\mib K}}

\def\bfa{{\mib a}}
\def\bfb{{\mib b}}

\def\bfd{{\mib d}}

\def\bfk{{\mib k}}

\def\bfn{{\mib n}}

\def\bfp{{\mib p}}

\def\bfz{{\mib z}}

\def\rbfn{{\bf n}}

\def\bfdelta{{\mib\delta}}
\def\eps{\epsilon}
\def\ve{\varepsilon}

\def\bfzeta{{\mib\zeta}}

\def\bftheta{{\mib\theta}}

\def\bfpi{{\mib\pi}}

\def\vphi{{\mib\varphi}}
\def\bfvphi{{\mib\vphi}}
\def\rvphi{{\raise.35ex\hbox{$\vphi$}}}
\def\rbvphi{{\raise.35ex\hbox{$\bfvphi$}}}

\def\xhi{{\raise.35ex\hbox{$\chi$}}}

\def\bfTheta{{\mib\Theta}}

\def\rmDelta{{\rm\Delta}}

\def\rmPsi{{\rm\Psi}}

\def\rmT{{\rm T}}

\def\rmc{{\rm c}}

\def\rmt{{\rm t}}

\def\xhat{{\hat{\mib x}}}
\def\yhat{{\hat{\mib y}}}
\def\zhat{{\hat{\mib z}}}

\def\nhat{{\hat{\mib n}}}

\def\bnabla{{\boldsymbol{\nabla}}}

\def\ket#1{{\big|\, #1 \big\rangle}}

\def\expect#1#2#3{{\big\langle #1 \,\big|\, #2 \big| #3 \big\rangle}}

\def\pz{{\partial}}

\def\nd{^{\vphantom{\dagger}}}
\def\ns{_{\vphantom{\dagger}}}
\def\yd{^\dagger}

\def\vph{\vphantom{\sum_i}}
\def\bvph{\vphantom{\sum_N^N}}

\def\and{a^{\phantom\dagger}}

\gdef\journal#1, #2, #3, 1#4#5#6{               
    {\sl #1~}{\bf #2}, #3 (1#4#5#6)}            

\physgreek

\begin{document}

\tolerance 10000

\def\yd{^\dagger}
\def\ve{\varepsilon}
\def\thbar{{\bar\theta}}
\def\vF{v_{\scriptscriptstyle\rm F}}
\def\BO{B\nd_{\rm \Omega}}
\def\Wpa{W\nd_\parallel}
\def\Wpe{W\nd_\perp}
\def\zhat{{\hat\bfz}}
\def\nhat{{\hat\bfn}}
\def\Sth{T(\theta\nd_1,\theta\nd_2)}
\def\Sths{T^*(\theta\nd_1,\theta\nd_2)}
\def\ns{^\vphantom{*}}
\def\bnabla{\mbox{\boldmath{$\twelvesy\nabla$}}}
\def\nhat{{\hat\rbfn}}
\def\uar{\uparrow}
\def\dar{\downarrow}
\def\cbar{{\bar c}}
\def\gze{\gamma\ns_0}
\def\gon{\gamma\ns_1}
\def\gtw{\gamma\ns_2}
\def\gth{\gamma\ns_3}
\def\gfo{\gamma\ns_4}
\def\gfi{\gamma\ns_5}
\def\psiC{\psi\nd_\ssr{C}}
\def\psiL{\psi\nd_\ssr{L}}
\def\psiR{\psi\nd_\ssr{R}}
\def\smi{\Sigma^-}
\def\spl{\Sigma^+}
\def\Mhat{{\widehat M}}
\def\ql{q\nd_{\ssr{L}\vphantom{\dagger}}}
\def\qr{q\nd_{\ssr{R}\vphantom{\dagger}}}
\def\xl{x\nd_{\ssr{L}\vphantom{\dagger}}}
\def\xr{x\nd_{\ssr{R}\vphantom{\dagger}}}
\def\yl{y\nd_{\ssr{L}\vphantom{\dagger}}}
\def\yr{y\nd_{\ssr{R}\vphantom{\dagger}}}
\def\al{\alpha\nd_{\ssr{L}\vphantom{\dagger}}}
\def\ar{\alpha\nd_{\ssr{R}\vphantom{dagger}}}
\def\bl{\beta\nd_{\ssr{L}\vphantom{\dagger}}}
\def\br{\beta\nd_{\ssr{R}\vphantom{\dagger}}}
\def\kl{\kappa\nd_{\ssr{L}\vphantom{\dagger}}}
\def\kr{\kappa\nd_{\ssr{R}\vphantom{\dagger}}}
\def\abar{{\bar\alpha}}
\def\bbar{{\bar\beta}}
\def\xbar{{\bar x}}
\def\ybar{{\bar y}}
\def\ns{^{\vphantom{*}}}
\def\ellb{\ell\ns_B}
\def\vf{v\ns_\ssr{F}}
\def\dbi{\rmDelta'}
\parskip=0pt

\title{Stacking Faults, Bound States, and Quantum Hall Plateaus in
Crystalline Graphite}

\author{ Daniel P. Arovas$^{1}$ and F. Guinea$^{2}$}

\affiliation{$^1$Department of Physics, University of California at
San Diego, La Jolla, CA 92093}
\affiliation{$^2$Instituto de Ciencia de Materiales de Madrid. CSIC. Cantoblanco.
E-28049 Madrid, Spain}

\begin{abstract}
We analyze the electronic properties of a simple stacking defect in
Bernal graphite.  We show that a bound state forms, which disperses
as $|\bfk-\bfK|^3$ in the vicinity of either of the two inequivalent
zone corners $\bfK$.  In the presence of a strong $c$-axis magnetic
field, this bound state develops a Landau level structure which for
low energies behaves as $E\nd_n\propto |n\,B|^{3/2}$.   We show that
buried stacking faults have observable consequences for surface
spectroscopy, and we discuss the implications for the
three-dimensional quantum Hall effect (3DQHE).  We also analyze the
Landau level structure  and chiral surface states of rhombohedral
graphite, and show that, when doped, it should exhibit multiple 3DQHE plateaus at
modest fields.
\end{abstract}

\date{\today}


\maketitle

\section{Introduction}\label{secint}
An explosion of research activity associated with the novel two-dimensional
material graphene has prompted a reexamination of its bulk parent, graphite.
Much, of course, is known about graphite \cite{BCP88}.  Bernal graphite is a hexgonal
crystal consisting of graphene sheets stacked in an ABAB configuration.
The $sp^2$-hybridized $\sigma$ electrons form double bonds between the carbon
atoms, while the remaining $\pi$ electrons, in the $p^z$ orbital, are itinerant.
The electronic structure parameters for graphite were first derived by Wallace, and
by Slonczewski, Weiss, and McClure (SWMC) \cite{SWMC}.
Within each plane, the $\pi$ electrons move on a honeycomb lattice with a nearest
neighbor hopping integral $\gze\approx 3.2\,$eV.  Of the four atoms
per unit cell, two are arranged in vertical chains, with a vertical nearest neighbor
hopping of $\gon\approx -390\,$meV.   Additional further neighbor hoppings are also
present.  For example, the $\pi$ electrons on the non-chain sites undergo
two-layer vertical hopping through open hexagons in the neighboring layers, with
amplitude $\half\gtw\approx -10\,$meV.  This
results in a very narrow band of width $40\,$meV along the $K$-$H$ spine of the
Brillouin zone, with electron pockets at $K$ and hole pockets at $H$ \cite{DM64}.

Recently, striking experimental observations of what may be bulk three-dimensional
quantum Hall plateaus in graphite has been reported \cite{KEK06}.   Any two-dimensional
(2D) system, such as graphene, which exhibits the quantum Hall effect (QHE) should
exhibit a 3DQHE if the interplane coupling is sufficiently weak.  The reason for this is
that the cyclotron gaps between Landau levels narrow continuously as one adiabatically
switches on the $c$-axis couplings, and cannot collapse immediately.  For a 3D electron
system in a periodic potential and subject to a magnetic field, a generalization of the
TKNN result \cite{TKNN} by Halperin \cite{Hal87} shows that the conductivity tensor must
be of the form
\begin{equation}
\sigma\nd_{ij} ={e^2\over h}\, \epsilon\nd_{ijk} \,G\nd_k\ ,
\end{equation}
whenever the Fermi level $E\nd_\ssr{F}$ lies within a bulk gap, where $\epsilon_{ijk}$
is the fully antisymmetric tensor and $\vec{G}$ is a reciprocal lattice vector of the
potential (which may be ${\vec G}=0$).  The Hall current is then carried by a sheath of
chiral surface states.  Eventually, however,
the $c$-axis hopping will become large enough that the Landau gaps collapse.
Equivalently, for a given value of the $c$-axis hopping $\gon$, the magnetic
field $B$ must exceed a critical strength $B\nd_\rmc$ in order that the Landau level
spacing overwhelms the $c$-axis bandwidth and opens up a bulk gap.

Typically, the field scale $B\nd_\rmc$ is extremely large, and much beyond the
scale of current experimentally available fields.
For a system with ballistic dispersion described by an effective mass
$m^*$, the orbital part of the spectrum (\ie\ neglecting Zeeman coupling)
yields a dispersion $\ve\nd_n=(n+\frac{1}{2})\,\hbar\omega\nd_\rmc$, where $n$
is a nonnegative integer and where $\omega\nd_\rmc=eB/m^* c$ is the cyclotron
frequency.  The cyclotron energy may be written as $\hbar\omega\nd_\rmc=
W\nd_\parallel\cdot(B/B\nd_\Omega)$ and the field scale as
$B\nd_\Omega =hc/e\Omega$, where $\Omega$ the unit cell area.  $W\nd_\parallel$ is
on the order of the bandwidth in zero field, which is typically several electron volts.
Since $hc/e=4.13\times 10^5\,{\rm T}{\rm\AA}^2$, $B\nd_\Omega$ is typically
enormous, on the order of tens of thousands of Tesla.  Thus, if the $c$-axis bandwidth
is $W\nd_\perp$, the critical field is given by
$B\nd_\rmc=(W\nd_\perp/W\nd_\parallel)\cdot B\nd_\Omega$, and even highly
anisotropic materials with $W\nd_\perp\ltwid 10^{-2}\,W\nd_\parallel$ will have
critical fields in the range of hundreds of Tesla.

As shown by Bernevig \etal\ \cite{BHRA07}, similar considerations would apply
for graphene sheets in AAA (simple hexagonal) stacking.   The Landau level dispersion
is then
\begin{equation}
E\nd_n(B,\bfk)=2\gon\cos(k\nd_z c)+{\rm sgn}(n)\,
\gze\sqrt{nB/B_0}\ ,
\end{equation}
with $B\nd_0=B\nd_\Omega\big/2\pi\sqrt{3}=7275\,{\rm T}$,
where $\gze\approx 3.16\,$eV is the in-plane hopping and $\gon\approx 0.39\,$eV
is the hopping integral between layers \cite{SWMC}.
The gap between Landau levels $n$ and $n+1$ collapses at a critical field
\begin{equation}
B\nd_{\rmc,n}=\bigg({4\gon\over\gze}\bigg)^{\!\!2}\cdot{B_0\over
\left(\sqrt{n+1}-\sqrt{n}\right)^2}\ .
\end{equation}
For $n=0$ one finds $B\nd_{\rmc,0}\approx 1800\,\rmT$.
However, due to the Bernal stacking, one finds \cite{BHRA07} that the principal
cyclotron gap surrounding the $n=0$ Landau levels opens above $B_\rmc=15\,\rmT$
(electrons; $n=+1$) or above $B_\rmc=7\,\rmT$ (holes; $n=-1$).  When the Fermi
level lies within either of these gaps, the Hall conductance is quantized at
$\sigma\nd_{xy}=2e^2/hd$, where $d=3.37\,$\AA\ is the inter-plane separation.

The analysis of ref. \cite{BHRA07} shows that the second cyclotron gap should not
open below fields on the order of $B\nd_{\rmc,2}\approx 1000\,\rmT$.  This suggests
that the multiple QHE plateaus observed by Kempa \etal\ \cite{KEK06} are of a different
origin, and are not describable by a model of crystalline Bernal graphite alone.

In this paper, we consider two variations which lead to a different plateau structure
to that of crystalline Bernal graphite.  The first is  rhombohedral graphite, which
is stacked in ABCABC fashion.  For this structure, we find
\begin{equation*}
B\nd_{\rmc,n}=\Big(\sqrt{n+1}+\sqrt{n}\Big)^{\!2}\,B_{\rmc,0}\ ,
\end{equation*}
with $B\nd_{\rmc,0}\approx 0.123\,\rmT$.  When $E\nd_\ssr{F}$ lies in the Fermi level
between the $n$ and $n+1$ Landau levels, the Hall conductivity is given by
$\sigma\nd_{xy}=(4n+2)\,e^2/h\, d$.  {\it Ab initio\/} calculations show
that the total energy of rhombohedral graphite to be approximately
$0.11\,$meV per atom larger than the Bernal hexagonal phase \cite{CGM94}.
With such a small energy difference, even highly oriented pyrolytic graphite (HOPG)
is believed to contain several percent rhombohedral inclusions.
Powdered graphite samples with up to $\sim 40\%$ of the rhombohedral phase
are obtainable \cite{CGAF00}.

The second possibility we examine is that of a simple stacking fault in Bernal
graphite, of the form ABABCBCB,  This fault interpolates between two
degenerate vacua -- the ABAB and CBCB Bernal phases.  We analyze the
$c$-axis transport through such a defect, within a simple model of nearest
neighbor hopping, and compute the $S$-matrix as a
function of in-plane wavevector.  As expected, the transmission is sharply
attenuated in the vicinity of the Dirac points.  We also find a novel bound state
associated with the stacking defect, with two-dimensional dispersion
$E(\bfk)\propto |\bfk-\bfK|^3$ near the Dirac points.  In the presence of a
$c$-axis magnetic field, this leads to a bound state Landau level energy
$E(n,B)\propto  |nB|^{3/2}$.  In the appendix, we undertake a calculation of
the bound state spectrum in zero field for the full SWMC model \cite{SWMC},
which includes seven tight binding parameters.

We conclude with a discussion of surface spectroscopy of buried stacking faults,
and with remarks about the relevance of our results to future experiments.

\begin{figure}[!t]
\centering
\includegraphics[width=8cm]{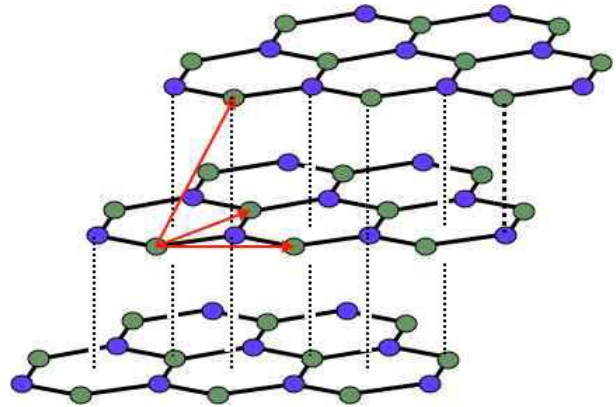}
\caption
{\label{RHG} Crystal structure of rhombohedral graphite.}
\end{figure}

\section{Rhombohedral Graphite}\label{rhombo}
In rhombohedral graphite (RG) there are two sublattices, in contrast to four in the case
of Bernal hexagonal graphite (BHG).  The primitive direct lattice vectors are
\begin{align*}
\bfa\nd_1&= a\,\xhat\\
\bfa\nd_2&=\half\, a\,\xhat + \frac{\sqrt{3}}{2}\,a\,\yhat\\
\bfa\nd_3&=\half\, a\,\xhat + \frac{1}{2\sqrt{3}}\,a\,\yhat + d\,\zhat \ .
\end{align*}
The basis vector $\bfdelta=-\frac{1}{3}\big(\bfa\nd_1+\bfa\nd_2\big)$
separates the $A$ and $B$ sublattices.  Note that $\bfa\nd_3=d\,\zhat -\bfdelta$.
The lattice parameters are $a=2.46\,$\AA\ and $d=3.37\,$\AA.

Our treatment starts with a simplified version of the work of McClure \cite{McC69}.
We consider several types of hopping processes:
\begin{itemize}
\item[(i)] in-plane $A-B$ hopping:
\begin{equation}
H_1^{AB}=-\gze\,\Big[ t(\bfdelta) + t(\bfa\nd_1+\bfdelta) + t(\bfa\nd_2+\bfdelta)\Big]\ ,
\end{equation}
where, $t(\bfd)$ is a translation operator through a vector $\bfd$.
\item[(ii)] neighboring plane diagonal $A-B$ hopping:
\begin{align}
H_2^{AB}&=\gth\Big[t(\bfa\nd_1-\bfa\nd_3+\bfdelta)+t(\bfa\nd_2-\bfa\nd_3+\bfdelta)\\
&\qquad\qquad\qquad+ t(\bfa\nd_1+\bfa\nd_2-\bfa\nd_3+\bfdelta)\Big]\,\nonumber
\end{align}
\item[(iii)] nearest neighbor and next nearest neighbor plane vertical $A-B$ hopping:
\begin{equation}
H_3^{AB}=\gon\,t(\bfa\nd_3+\bfdelta)+\gtw\,t(\bfa\nd_1+\bfa\nd_2-2\bfa\nd_3+\bfdelta)
\end{equation}
\item[(iv)] neighboring plane diagonal $A-A$ hopping:
\begin{equation}
H_4^{AA}=\gth\Big[t(\bfa\nd_3)+t(\bfa\nd_3-\bfa\nd_1)
+t(\bfa\nd_3-\bfa\nd_2)\Big] + {\rm H.c.}
\end{equation}
\item[(v)] neighboring plane diagonal $B-B$ hopping:
\begin{equation}
H_4^{BB}=\gth\Big[t(\bfa\nd_3)+t(\bfa\nd_3-\bfa\nd_1)
+t(\bfa\nd_3-\bfa\nd_2)\Big]+ {\rm H.c.}
\end{equation}
\end{itemize}
The full Hamiltonian is then given by
\begin{equation}
H=\begin{pmatrix} H_4^{AA} & H_1^{AB}+H_2^{AB}+H_3^{AB} \\ &&\\
 H_1^{BA}+H_2^{BA}+H_3^{BA}  &H_4^{BB}\end{pmatrix}\ ,
\end{equation}
where $H_n^{BA}=\big(H_n^{AB}\big)\yd$ for $n=1,2,3$.
From Wallace and SWMC \cite{SWMC}, we take
\begin{align*}
\gze&=3160\,{\rm meV}\qquad&\gon&=390\,{\rm meV}\\
\gtw&=10\,{\rm meV}
\qquad&\gth&=250\,{\rm meV}\ .
\end{align*}
(In the language of McClure \cite{McC69}, $\gamma'_2=\gtw$ and $\gamma'_1=\gth$,
and we ignore McClure's parameters $\gamma'_0$ and $\gamma''_2$.)
We then have
\begin{equation}
H=\begin{pmatrix} \eta & 0 \\ 0 & 1\end{pmatrix}
\, \begin{pmatrix} A & B \\ B^* & A \end{pmatrix}\,
\begin{pmatrix} \eta^* & 0 \\ 0 & 1\end{pmatrix}\  ,
\end{equation}
with $\eta= e^{i(\theta\nd_1+\theta\nd_2)/3}$ and
\begin{align*}
A(\theta\nd_1,\theta\nd_2,\theta\nd_3)&=\gth\,e^{-i\theta\nd_3}\,\Sth +\gth\,
e^{i\theta\nd_3}\,\Sths\\
B(\theta\nd_1,\theta\nd_2,\theta\nd_3)&=-\gze\, \Sth +\gth\,e^{-i\theta\nd_3}\,\Sths\\
&\qquad\qquad +\gon\,e^{i\theta\nd_3} +\gtw\,e^{i(\theta\nd_1+\theta\nd_2-2\theta\nd_3)}
\end{align*}
where
\begin{equation}
\Sth=1+e^{i\theta\nd_1} + e^{i\theta\nd_2} \ .
\end{equation}
The energy eigenvalues are clearly
\begin{equation}
E\nd_\pm(\bftheta)=A(\bftheta)\pm \big| B(\bftheta)\big|\ .
\end{equation}
Under a $60^\circ$ rotation, we have
\begin{equation}
\theta'_1=\theta\nd_2\quad,\quad\theta'_2=\theta\nd_2-\theta\nd_1
\quad,\quad\theta'_3=\theta\nd_2-\theta\nd_3\ .
\end{equation}
One then finds $A(\bftheta')=A(\bftheta)$ and $B(\bftheta')=e^{i\theta\nd_2}\,
B(\bftheta)$.  Hence, $E\nd_\pm(\bftheta')=E\nd_\pm(\bftheta)$.

\begin{figure}[!t]
\centering
\includegraphics[width=6cm, height=7cm]{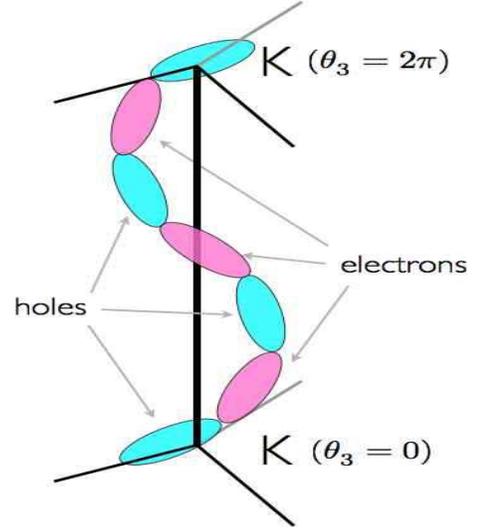}
\caption
{\label{FS} McClure's ``sausage link" Fermi surface for rhombohedral graphite, greatly
exaggerated.  See also fig. 2 of ref. \cite{McC69}.}
\end{figure}

Degeneracies identified with a one-parameter family of Dirac points occur when
$B(\bftheta)=0$. Solving, we obtain the relation
\begin{equation}
\Sth=\Gamma\nd_1\,e^{i\theta\nd_3} +
\Gamma\nd_2\,e^{i(\theta\nd_1+\theta\nd_2-2\theta\nd_3)}
\label{SDir}
\end{equation}
along the degeneracy curve, where
\begin{align}
\Gamma\nd_1&\equiv{\gze\,\gon+\gtw\,\gth \over\gamma_0^2 -\gamma_3^2}=-0.124\\
\bvph \Gamma\nd_2&\equiv {\gon\,\gth+\gze\,
\gtw \over\gamma_0^2 -\gamma_3^2}=-1.30\times 10^{-2}\ .
\end{align}
The energy along this Dirac curve is
\begin{equation}
E(\bftheta\nd_\ssr{D})=\cE\nd_0 +W\,
\cos\big(\theta\nd_1+\theta\nd_2 - 3\theta\nd_3\big)\ .
\label{ADir}
\end{equation}
with
\begin{align}
\cE\nd_0&=2\Gamma\nd_1\,\gth=62\,{\rm meV}\\
W&=2\Gamma\nd_2\,\gth=6.5\,{\rm meV} .
\end{align}
Since $\Gamma\ns_1$ and $\Gamma\ns_2$ are small, the Dirac curve, when projected
into the basal Brillouin zone, lies close to the zone corners.  Note that
$E(\bftheta\nd_D)$ goes through three complete periods
as $\theta\nd_3$ advances from $0$ to $2\pi$, resulting in McClure's `sausage link'
Fermi surface \cite{McC69}, depicted in fig. \ref{FS}.  To find the equation of the Dirac
curve, we expand about $\bfTheta=(\theta\nd_1,\theta\nd_2)=
\big(\frac{4\pi}{3},\frac{2\pi}{3}\big)$ at the $K$ point,
writing $\bftheta=\bfTheta\nd+\bfzeta$, and find
\begin{equation}
T\big(\Theta\nd_1+\delta\theta\nd_1,\Theta\nd_2+\delta\theta\nd_2)=e^{-i\pi/6}\,\delta\theta\nd_1
-e^{i\pi/6}\,\delta\theta\nd_2 + \cO(\delta\theta^2)\ .
\label{Sexp}
\end{equation}
Solving for the Dirac line $\bfzeta(\theta\nd_3)$ as a formal series in the
small parameters $\Gamma\nd_1$ and $\Gamma\nd_2$, we obtain
\begin{align*}
\delta\theta\nd_1&=\frac{2}{\sqrt{3}}\,\Big[\!-\!\Gamma\nd_1\,
\sin\big(\theta\nd_3-\frac{\pi}{6}\big)
+\Gamma\nd_2\,\sin\big(2\theta\nd_3+\frac{\pi}{6}\big)\Big]+\cO(\Gamma^2)\\
\delta\theta\nd_2&=\frac{2}{\sqrt{3}}\,\Big[\Gamma\nd_1\,\sin\big(\theta\nd_3+\frac{\pi}{6}\big)
-\Gamma\nd_2\,\sin\big(2\theta\nd_3-\frac{\pi}{6}\big)\Big]+
\cO(\Gamma^2)\ .
\end{align*}

Note that the bandwidth of the Dirac point energies is tiny:
$2W\approx 13\,{\rm meV}$.
This means that the Landau levels are quite narrow -- moreso than in Bernal
stacked graphite.
The Fermi surface resembles the sketch in fig. \ref{FS}, which is adapted from fig. 2
of ref. \cite{McC69}

\subsection{Weak Fields : Kohn-Luttinger Substitution}\label{weak}
We assume the magnetic field $\bfB$ is directed along $\zhat$.
To obtain the Landau levels, we expand about the Dirac points.  (This is essentially
equivalent to expanding about the Fermi energy, since the bandwidth of the Dirac
points is so tiny.)  We write
\begin{equation}
\bfk\longrightarrow\bfK + \hbar^{-1}\,\bfpi\ ,
\end{equation}
where $\bfpi=\bfp+\frac{e}{c}\bfA$.  With $\delta\theta\nd_j=(\bfk-\bfK)\cdot\bfa\nd_j$,
we have
\begin{align}
\delta\theta\nd_1&={1\over\hbar}\,\pi\nd_x\,a\\
\delta\theta\nd_2&={1\over 2\hbar}\,\pi\nd_x\,a + {\sqrt{3}\over 2\hbar}\,\pi\nd_y\,a\ .
\end{align}
Recall $[\,\pi\nd_x,\pi\nd_y\,]=-i\hbar^2/\ell_B^2$ where $\ellb=\sqrt{\hbar c/eB}$ is the
magnetic length.  From eqn. \ref{Sexp},  to lowest order in $\delta\theta\nd_{1,2}$, we have
\begin{align}
\big[\, T \, , \, T\yd \, \big]&=2i\sin(\pi/3)\,
\big[\,\delta\theta\nd_1 \, , \, \delta\theta\nd_2\,\big]\nonumber\\
&=2\pi\sqrt{3}\,p/q\ .
\end{align}
where the flux per unit cell area $\Omega=\frac{\sqrt{3}}{2}\,a^2$ is assumed to be a
rational multiple $p/q$ of the Dirac flux quantum $\phi\nd_0=hc/e$.
This means we may write
\begin{equation}
\Sth=-\,\eps\,b\ ,
\end{equation}
where
\begin{equation}
\eps=\sqrt{B/B_0}\ ,
\label{epse}
\end{equation}
and $b\yd$ is a Landau level raising operator: $\big[b,b\yd]=1$.  Recall that the field
scale $B\nd_0=(hc/e)/3\pi a^2=7275\,\rmT$.
It is convenient to define $\thbar\ns_3=\theta\ns_3-\frac{1}{3}(\theta\ns_1+\theta\ns_2)$,
and to absorb a phase into the definition of $b$,
taking $T=-\,\eps\,e^{-i\thbar_3}\,b\yd$.  Note that when the magnetic field lies
along the $c$-axis, it is $\exp(i\thbar\ns_3)$ and not
$\exp(i\theta\ns_3)$ which commutes with the magnetic translations $t(\bfa\ns_{1,2})$.
The Hamiltonian is then
\begin{align}
H&=\cE\nd_0 + W\cos(3\thbar\nd_3) \\
&\qquad + \eps
\begin{pmatrix} -\gth \,(b +b\yd) &
\gze\,e^{-i\thbar_3}\,b\yd -\gth\,b\\
\gze\,e^{i\thbar_3}\,b -\gth\,b\yd &
-\gth \,(b + b\yd) \end{pmatrix}\ .
\end{align}

Consider the matrix operators
\begin{align}
\cQ\nd_0&=\gze\begin{pmatrix} 0 & e^{-i\thbar_3}\,b\yd \\
e^{i\thbar_3}\,b & 0 \end{pmatrix} \\
\cQ\nd_1&=\gth\begin{pmatrix}  b +b\yd & b\\ b\yd & b + b\yd
\end{pmatrix}
\end{align}
The eigenvectors of $\cQ\nd_0$ are
\begin{equation}
\ket{\rmPsi\nd_0}=\begin{pmatrix} \ket{0} \\  0 \end{pmatrix} \quad,\quad
E^0_0=0
\end{equation}
and
\begin{equation}
\ket{\rmPsi^\pm_n}={1\over\sqrt{2}}\begin{pmatrix} e^{-i\thbar_3}\,\ket{n} \\
\pm\, \ket{n-1} \end{pmatrix} \quad,\quad
E^0_n=\pm\, \sqrt{n}\,\eps\,\gze\ ,
\end{equation}
where $n=1,2,3,\ldots\ .$ It is easy to see that
\begin{equation}
\expect{\rmPsi^\pm_n}{\cQ\nd_1}{\rmPsi^\pm_n}=0\ ,
\end{equation}
as well as $\expect{\rmPsi\nd_0}{\cQ\nd_1}{\rmPsi\nd_0}=0$, hence there is no first
order shift of the eigenvalues.  Therefore, up to first order in $\eps$, the Landau
level energies are given by
\begin{equation}
E\nd_n(\theta\nd_3)=\cE\nd_0+W\cos(3\thbar\nd_3)\pm\eps\,\gze\,\sqrt{n}\ ,
\end{equation}
where $n=0,1,2,\ldots\ .$  The gap between Landau levels $n$ and $n+1$ collapses when
\begin{equation}
\eps\,\gze\,\sqrt{n}+W=\eps\,\gze\,\sqrt{n+1}-W\ ,
\end{equation}
which gives a critical field of
\begin{equation}
B\nd_{\rmc,n}=\Big(\sqrt{n+1}+\sqrt{n}\Big)^{\!2}\,B_{\rmc,0}\ ,
\end{equation}
with $B\nd_{\rmc,0}=(2W/\gze)^2\cdot B\nd_0=0.123\,$\rmT.

\begin{figure}[!b]
\centering
\includegraphics[width=7cm]{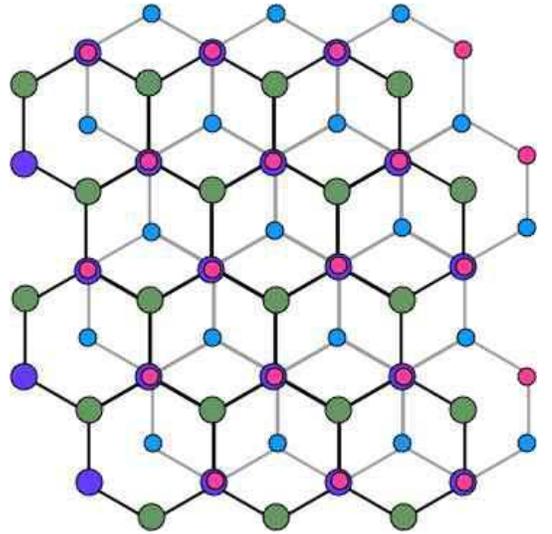}
\caption
{\label{bernal_2} Top-view of Bernal hexagonal graphite.}
\end{figure}

\subsection{Comparison with Bernal Stacking}
The $ABAB$ stacking pattern of Bernal hexagonal graphite is shown in fig. \ref{bernal_2}.
To obtain the critical fields in BHG, it suffices to consider
a simple nearest-neighbor model \cite{BHRA07}.  Expanding about the $K$-$H$ spine
in the Brillouin zone, we obtain in the presence of a uniform $c$-axis magnetic field,
\begin{equation}
H = \left(
\begin{array}{cccc}
  0 & \eps\,\gze\, b & 2\,\gon \cos\theta\nd_3 & 0 \\
\eps\,t\nd_\parallel\, b\yd & 0 & 0 & 0 \\
  2\, \gon \cos\theta\nd_3 & 0 & 0 & -\eps\,\gze\,  b\yd \\
  0 & 0 & \eps\,\gze\, b & 0 \\
\end{array}\right) \label{toy}
\end{equation}
where $\eps=(2\pi\sqrt{3}\,p/q)^{1/2}=\sqrt{B/B_0}$ as in the rhombohedral case.
The spectrum has explicit particle-hole symmetry. For $n=0$ there
are eigenvalues at
$\pm\big(\eps^2\,\gamma_0^2 + 4\gamma_1^2\,\cos^2\!\theta\nd_3\big)^{1/2}$
and a doubly degenerate level at $E\nd_0=0$.  For $n\ne 0$,
\begin{align}
E^2_n=&(n+\half)\, \eps^2\, \gamma_0^2+2\gamma_1^2
\cos^2\!\theta\nd_3\\
&\hskip -0.6cm \pm\sqrt{\frac{1}{4}\,\eps^4\,\gamma_0^4+ 4\,(n+\half) \,\eps^2\,
\gamma_0^2\,\gamma_1^2\, \cos^2\!\theta\nd_3+
4\, \gamma_1^4\,\cos^4\!\theta\nd_3} \ .\nonumber
\end{align}

In fig. \ref{BHG_bands}, we plot the lowest several energy bands {\it versus\/} magnetic
field for BHG.  Due to the inadequacies of the nearest neighbor model, the principal
gap surrounding central $E=0$ Landau levels opens immediately for nonzero $B$.
Including more realistic band structure effects, consistent with the semimetallic nature
of BHG, this gap opens at a critical field of $B\nd_\rmc\approx 15\,\rmT$ for positive
energies and $B\nd_\rmc\approx 7\,\rmT$ for negative energies \cite{BHRA07}.
The Hall conductance
is quantized at $\sigma\nd_{xy}=2C\,e^2/hc\nd_0$ when the Fermi level lies in a bulk gap,
where $c\nd_0=2d$ in BHG and $c\nd_0=3d$ in RG, where $d=3.37\,$\AA\ is the spacing
between planes, and $C$ is a topological integer associated with the gap.
In both cases, the values of $C$ are such that $\sigma\nd_{xy}$ corresponds to
the graphene quantization per layer, changing by $4 e^2/hd$ as one crosses a Landau
level.  We indicate the width of the bands by shading
the region between $\cos^2\!\theta\nd_0=0$ and $\cos^2\!\theta\nd_3=1$.
In both cases, the Zeeman coupling is omitted; with $g\approx 2$ the Zeeman
splitting is small compared with the cyclotron energy.

\begin{figure}[!t]
\centering
\includegraphics[width=8cm]{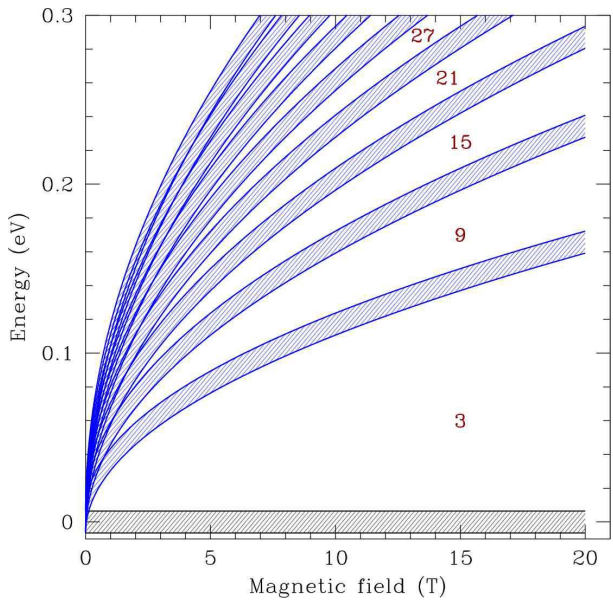}
\caption
{\label{RHG_bands} Landau level structure in rhombohedral graphite within the
tight binding model of section \ref{rhombo}, with Zeeman term ignored.  Principal band
gaps are labeled by the Chern number $C$ (per spin degree of freedom).  When $E_\ssr{F}$
lies within a gap, the Hall conductivity is $2C\times\frac{e^2}{h}\big/(3d)$, where $d=3.37\,$
\AA\ is the interplane spacing.}
\end{figure}

\begin{figure}[!t]
\centering
\includegraphics[width=8cm]{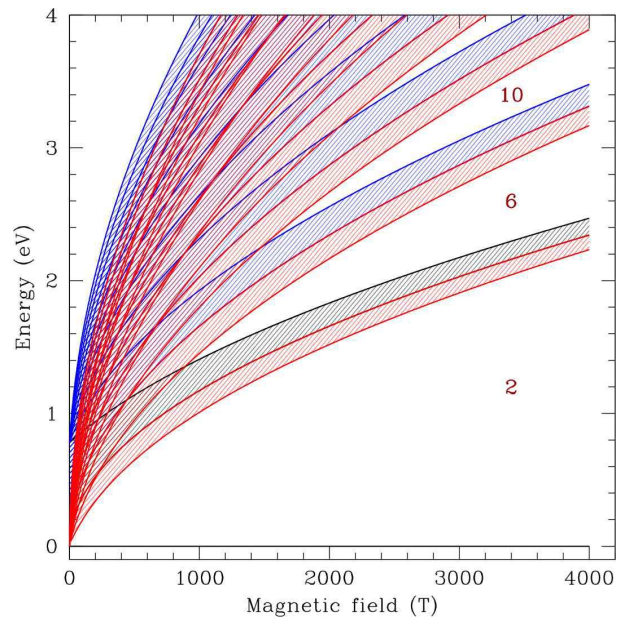}
\caption
{\label{BHG_bands} Landau level structure in Bernal graphite within the nearest-neighbor
hopping model, with Zeeman term ignored.  Principal band gaps are labeled by the
Chern number $C$ (per spin degree of freedom).  When $E_\ssr{F}$ lies within a gap,
the Hall conductivity is $2C\times \frac{e^2}{h}\big/(2d)$.  When further neighbor hoppings are
included, particle-hole symmetry is
broken, and a finite field is required to open the principal gap. \cite{BHRA07}.}
\end{figure}

\section{Chiral Surface States}
As shown by Hatsugai \cite{Hat93}, the Chern number $C$ can also be computed by
following the spectral flow in a system with edges, wrapped around a cylinder, as
a function of the gauge flux through the cylinder.  To elicit this spectral flow, we
derive a Hofstadter Hamiltonian \cite{Hof76} for RG.  We start with the Hamiltonian
elements in section \ref{rhombo}, but now treating them as magnetic translations,
which satisfy the algebra
\begin{equation}
t(\bfa)\,t(\bfb)=e^{i\bfa\times\bfb\cdot\nhat/2\ell_B^2}\,t(\bfa+\bfb)\ ,
\end{equation}
where $\bfB=B\,\nhat$.  For our problem we define the elementary translations
\begin{equation}
t\nd_1\equiv t(\bfdelta) \quad,\quad t\nd_2\equiv t(\bfa\nd_1+\bfdelta)
\end{equation}
as well as $\tau\equiv t(d\zhat)=t(\bfa\nd_3+\bfdelta)$.

\begin{figure}[!t]
\centering
\includegraphics[width=8cm]{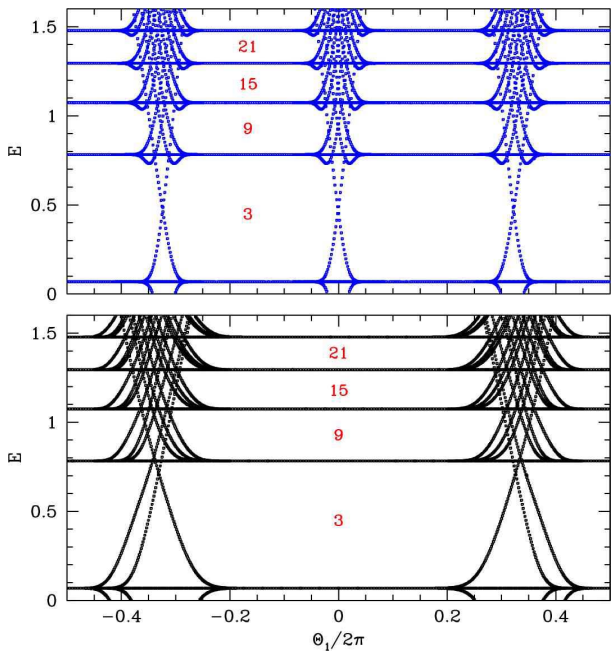}
\caption
{\label{RHG_edge} Spectral flow in rhombohedral graphite showing edge state evolution.
Top panel: armchair edge, perpendicular to $\bfdelta$; bottom panel: zigzag edge,
perpendicular to $\bfa_1$.  The bulk gaps are labeled by Chern numbers $C$ which
correspond to the number of edge states crossing the gap as the angle $\theta_1$
is varied.  The flux per unit cell here is rather large, with $p=1$ and $q=200$,
corresponding to a field of $B=396\,\rmT$.  The topological features of the
edge state spectral flow are robust with respect to field.}
\end{figure}

Since with $\nhat=\zhat$ we have that $\tau$ commutes with $t\nd_1$ and $t\nd_2$,
and we can specify its eigenvalue as $e^{i\thbar\nd_3}$.  As for $t\nd_{1,2}$,
we have
\begin{equation}
t\nd_1\>t\nd_2=e^{i\phi/3}\,t\nd_2\>t\nd_1\ ,
\end{equation}
where $\phi=\Omega/\ell_B^2=2\pi p/q$ is the flux per graphene hexagon in units
of $\hbar c/e$.  We may then write
\begin{equation}
H=\begin{pmatrix} H\nd_{AA} & H\nd_{AB} \\ & \\ H\yd_{AB} & H\nd_{BB} \end{pmatrix}\ ,
\end{equation}
with
\begin{align}
H\nd_{AA}&=\gth\,\big(e^{i\thbar_3}\,t\nd_1 + e^{-i\thbar_3}\,
t_1\yd\big)\\
&\qquad + \gth\,\big(e^{i\thbar_3} + e^{-i\thbar_3}\,
e^{-i\phi/6}\,t\nd_1\big)\,t\nd_2\nonumber\\
&\qquad + \gth\,\big(e^{-i\thbar_3} + e^{i\thbar_3}\,
e^{-i\phi/6}\,t_1\yd\big)\,t_2\yd=H\nd_{BB}\nonumber
\end{align}
and
\begin{align}
H\nd_{AB}&=\big(\gon\,e^{i\thbar_3}+\gtw\,e^{-2i\thbar_3}
-\gze\,t\yd_1 + \gth\,e^{-i\thbar_3}\,t\nd_1\big)\nonumber\\
&\qquad\qquad  +\big(\gth\,e^{-i\thbar_3} - \gze\,e^{-i\phi/6}\,t\nd_1\big)\,t\nd_2
\nonumber\\
&\qquad\qquad -\big(\gze - \gth\,e^{i\thbar_3}\,e^{-i\phi/6}\,t\yd_1\big)\,t\yd_2\ .
\end{align}
We define the basis $\big\{\ket{n}\big\}$ as follows:
\begin{align}
t\nd_1\,\ket{n}&=e^{i{\bar\theta}\nd_1}\,e^{i n\phi/3}\,\ket{n} \\
t\nd_2\,\ket{n}&= e^{i\theta\nd_2}\,\ket{n+1}\ ,
\end{align}
where ${\bar\theta}\nd_1=\theta\nd_1/3q$ and $\ket{n+3q}=\ket{n}$.
Taking the matrix elements of $H$ within this basis, one obtains a rank $6q$ matrix
$H$ to diagonalize, with periodic boundary conditions.  If we introduce an edge
by eliminating the coupling between states $\ket{1}$ and $\ket{3q}$, and plot the
spectral flow as a function of ${\bar\theta}\nd_1$, we obtain the top panel in fig.
\ref{RHG_edge}.  We can also obtain the chiral surface state flow for a zigzag edge,
perpendicular to the vector $\bfa\nd_1$; this is shown in the bottom panel of
fig. \ref{RHG_edge}.  For periodic systems, exact diagonalizations performed using the
Lanczos method for $q$ up to $1500$ with the package ARPACK were found to agree with
the weak field results of section \ref{weak}.

\section{Stacking Faults in Bernal Hexagonal Graphite}
We now turn to an analysis of simple stacking faults in BHG, first with $\bfB=0$ and
then for finite $\bfB$.  Consider first a triangular lattice, which is tripartite, and its
three triangular sublattices $A$, $B$, and $C$.  By eliminating one of these three
sublattices, the remaining structure will be a honeycomb lattice.  Now imagine
a stack of triangular lattices.  At each layer, we choose a sublattice $A$, $B$, or $C$
to remove; this defines a stacking pattern.  Since it is energetically unfavorable to
stack a honeycomb layer directly atop another, at each layer we have two choices
consistent with the layer below.  If the empty sublattices are in $ABC$ {\it et cyc.\/}
order from layer $l$ to layer $l+1$, we write $\sigma\nd_{n,n+1}=+1$.  If instead the
order is $CBA$ {\it et cyc.\/}, we write $\sigma\nd_{l,l+1}=-1$.  For RG, the $\sigma$
indices are `ferromagnetic', \ie\ $++++$ or $----$.  For BHG, the indices are ordered
`antiferromagnetically', \ie\ $+-+-$.

The three three triangular sublattices A, B, and C are defined by
\begin{align}
u^\ssr{A}_{n\ns_1,n\ns_2}&=n\ns_1\,\bfa\ns_1 + n\ns_2\, \bfa\ns_2\\
u^\ssr{B}_{n\ns_1,n\ns_2}&=n\ns_1\,\bfa\ns_1 + n\ns_2\, \bfa\ns_2 + \bfdelta\ns_1\\
u^\ssr{C}_{n\ns_1,n\ns_2}&=n\ns_1\,\bfa\ns_1 + n\ns_2\, \bfa\ns_2 + 2\,\bfdelta\ns_1
\end{align}
We define three additional sublattices by
\begin{align}
v^\ssr{A}_{n\ns_1,n\ns_2}&=u^\ssr{B}_{n\ns_1,n\ns_2}=n\ns_1\,\bfa\ns_1 + n\ns_2\, \bfa\ns_2 + \bfdelta\ns_1\\
v^\ssr{B}_{n\ns_1,n\ns_2}&=u^\ssr{C}_{n\ns_1,n\ns_2}=n\ns_1\,\bfa\ns_1 + n\ns_2\, \bfa\ns_2 + 2\,\bfdelta\ns_1 \\
v^\ssr{C}_{n\ns_1,n\ns_2}&=u^\ssr{A}_{n\ns_1,n\ns_2}=n\ns_1\,\bfa\ns_1 + n\ns_2\, \bfa\ns_2
\end{align}
The sites $\big\{u\ns_\ssr{A}(n\ns_1,n\ns_2)\, , \, v\ns_\ssr{A}(n\ns_1,n\ns_2)\big\}$ \etc\ form a honeycomb lattice, which we call
the $A$ or $\alpha$ structure.  Bernal graphite is stacked in an $ABABAB$ configuration.

Within each honeycomb layer, we write the wavefunction as a two-component spinor,
\begin{equation}
\psi\nd_\bfk=\begin{pmatrix} u\nd_\bfk \\ v\nd_\bfk \end{pmatrix}\ ,
\end{equation}
where $\bfk$ is the crystal momentum in the basal ($k_z=0$) Brillouin zone.

The hopping between planes is described by the following local Schr{\"o}dinger equation,
which couples a central plane $l$ to planes below ($l-1$) and above ($l+1$):
\begin{equation}
M\psi\ns_l+\gon(\Sigma^{\sigma})\yd\psi\ns_{l-1} +
\gon\Sigma^{\sigma'}\psi\ns_{l+1}=0\ .
\label{PSE}
\end{equation}
Here, $\sigma\ns_{l-1,l}=\sigma$ and $\sigma\ns_{l,l+1}=\sigma'$,
\ie\ the shift in the $u$ sublattice sites from plane $l-1$ to plane $l$ is through
a vector $\sigma\bfdelta\ns_1$.  The matrix $M$ is given by
\begin{equation}
M=\begin{pmatrix} E & \gze\,S \\ \gze\,S^* & E \end{pmatrix}\ ,
\end{equation}
and
\begin{equation}
S=e^{i\bfk\cdot\bfdelta\nd_1} + e^{i\bfk\cdot\bfdelta\nd_2} +
e^{i\bfk\cdot\bfdelta\nd_3}
\end{equation}
and
\begin{equation}
\Sigma^+=\begin{pmatrix} 0 & 1 \\ 0 & 0 \end{pmatrix} \qquad,\qquad
\Sigma^-=\begin{pmatrix} 0 & 0 \\ 1 & 0 \end{pmatrix}\ .
\end{equation}

\subsection{Bernal Hexagonal Graphite}
We first consider the BHG stacking order $ABABAB$, where $\sigma\ns_{l,l+1}=(-1)^l$.
Using translational invariance, we may write, for the even and odd sites
\begin{align}
\psi\nd_{2j}&=e^{2ijk_z d}\,\phi\\
\psi\nd_{2j+1}&=e^{i(2j+1)k_z d}\,\xhi\ ,
\end{align}
where
\begin{align}
M\phi +2\gon\cos\big(k\nd_z d\big)\smi\xhi&=0\vph\\
M\xhi +2\gon\cos\big(k\nd_z d\big)\spl\phi&=0\ .
\end{align}
Inverting the second of these equations gives
\begin{equation}
\xhi=-2\gon\cos\big(k\nd_z d\big) M^{-1}\spl\phi\ .
\label{invert}
\end{equation}
Substituting this into the first equation yields
\begin{equation}
\Big(M-4\gamma_1^2\cos^2\!\big(k\nd_z d\big) \smi M^{-1}\spl\Big)\phi=0\ .
\end{equation}
Accordingly, we define
\begin{align}
\cK&\equiv M-4\gamma_1^2\cos^2\!\big(k\nd_z d\big) \smi M^{-1}\spl\vph\\
&=\begin{pmatrix} E & \gze\,S \\ & \\ \gze\,S^* & E\Big(1-{4\gamma_1^2
\cos^2\!(k\nd_z d )\over E^2-\gamma_0^2\, |S|^2}\Big)\end{pmatrix}\ .
\end{align}
Setting ${\rm det}\,\cK=0$ yields the eigenvalue equation for Bernal graphite,
\begin{equation}
\big(E^2-\gamma_0^2\, |S|^2\big)^2-4E^2\,\gamma_1^2
\cos^2\!\big(k\nd_z d\big)=0\ ,
\end{equation}
with solutions
\begin{equation}
E^{(\mu,\mu')}_{\bfk,k\nd_z}=-\mu\,\gon\cos\big(k\nd_z d\big)
-\mu'\sqrt{\gamma_1^2\cos^2\!\big(k\nd_z d\big) + \gamma_0^2\,|S|^2}\ ,
\end{equation}
where $\mu=\pm 1$ and $\mu'=\pm 1$.  The four choices for $(\mu,\mu')$ correspond to the
four energy bands.

From $\cK\,\phi=0$, we may write
\begin{equation}
\phi=\begin{pmatrix}\phi\nd_1 \\ \phi\nd_2\end{pmatrix}=
\begin{pmatrix} -\gze\,S \\ E\end{pmatrix}\ .
\end{equation}
From eqn. \ref{invert}, then, we have
\begin{equation}
\xhi=\begin{pmatrix}\xhi\nd_1 \\ \xhi\nd_2\end{pmatrix}=
{2E\,\gon\cos(k\nd_z d )\over E^2-\gamma_0^2\, |S|^2}\,
\begin{pmatrix} -E \\ \gze\,S^* \\ \end{pmatrix}=\mu\begin{pmatrix}
E \\ -\gze\,S^*\end{pmatrix}\ .
\end{equation}

\subsection{Step Defect}
Consider now the stacking defect $ABABCBCB$, which in terms of the
$\sigma\ns_{l,l+1}$ variables may be depicted as
\begin{equation}
\cdots|+|-|+|-|-|+|-|+|\cdots
\label{step}
\end{equation}
The central plane we label  $l=0$.  For plane indices $l<0$, the odd layers correpond to $\phi$
planes and the even layers to $\xhi$ planes.  For $l>0$, the even layers correspond to
$\phi$ planes and the odd layers to $\xhi$ planes.  With $l<0$, we consider an incident
plane wave running to the right (up) and a reflected plane wave running to the left
(down).  Then we have
\begin{align}
\psi\nd_{2j}&=\big(\alpha\,e^{2ijk\nd_z d} + \beta'\, e^{-2ijk\nd_z d} \big)\,\xhi\\
\psi\nd_{2j+1}&=\big(\alpha\,e^{i(2j+1)k\nd_z d} + \beta'\,
e^{-i(2j+1)k\nd_z d} \big)\,\phi\ ,
\end{align}
for all $j<0$.  Here $\alpha$ is the complex amplitude of the incident wave and
$\beta'$ is the complex amplitude of the reflected wave.

Correspondingly, we have
\begin{align}
\psi\nd_{2j-1}&=\big(\beta\,e^{i(2j-1) k\nd_z d}+\alpha' \,
e^{-i(2j-1) k\nd_z d}\big)\,\xhi\\
\psi\nd_{2j}&=\big(\beta\, e^{2ijk\nd_z d}+\alpha'\,e^{-2ijk\nd_z d}\big) \,\phi\ ,
\end{align}
for all $j>0$.  Here $\alpha'$ is the incident amplitude (from the right/top)
and $\beta$ is the reflected amplitude.

To match the solutions for positive and negative $l$, we first invoke eqn. \ref{PSE} with $l=-1$:
\begin{equation}
M\psi\nd_{-1} + \gon\smi\psi\nd_{-2} + \gon\smi\psi\nd_0=0\ .
\end{equation}
The most general solution for $\psi\nd_0$ is then
\begin{equation}
\psi\nd_0=(\alpha+\beta')\,\xhi + \begin{pmatrix} 0 \\ b \end{pmatrix}\ ,
\label{left}
\end{equation}
where $b$ is an arbitrary complex number.  Note that $\smi$ annihilates any
vector with upper component $0$.

Next, set $l=+1$ and obtain
\begin{equation}
M\psi\nd_{1} + \gon\spl\psi\nd_{0} + \gon\spl\psi\nd_2=0\ .
\end{equation}
We may now write
\begin{equation}
\psi\nd_0=(\beta+\alpha')\,\phi + \begin{pmatrix} a \\ 0 \end{pmatrix}\ ,
\label{right}
\end{equation}
where $a$ is an arbitrary complex parameter.  Note that $\spl$ annihilates any vector
with lower component $0$.

The parameters $a$ and $b$ are then fixed by equating these two expressions for
$\psi\nd_0$, yielding
\begin{equation}
\begin{pmatrix} a \\ -b \end{pmatrix} = (\alpha+\beta')\,\xhi - (\beta+\alpha') \,\phi\ .
\end{equation}
The wavefunction at $l=0$ can now be found.  One simple way is to take the upper
component from eqn. \ref{left} and the lower component from eqn. \ref{right}:
\begin{equation}
\psi\nd_0=\begin{pmatrix} (\alpha+\beta')\,\xhi\nd_1 \\ (\beta+\alpha')\,\phi\nd_2
\end{pmatrix}\ .
\end{equation}

Next, we write the Schr{\"o}dinger equation for the $l=0$ plane:
\begin{align}
0&=M\psi\nd_0 + \gon\spl\psi\nd_{-1} + \gon\smi\psi\nd_{+1}\\
&=\begin{pmatrix} E & \gze\,S \\ \gze S^* & E \end{pmatrix}
\begin{pmatrix} (\alpha+\beta')\,\xhi\nd_1 \\ (\beta+\alpha')\,\phi\nd_2 \end{pmatrix}
\nonumber\\
&\qquad\qquad+\gon\,\big(\alpha\,e^{-i k\nd_z d} + \beta'\,
e^{i k\nd_z d}\big)
\begin{pmatrix} \phi\nd_2 \\ 0 \end{pmatrix}\nonumber\\
&\qquad\qquad\qquad +
\gon\,\big(\beta\,e^{i k\nd_z d} + \alpha'\, e^{-i k\nd_z d}\big)\,
\begin{pmatrix} 0 \\ \xhi\nd_1 \end{pmatrix}\nonumber\ .
\end{align}
This yields two equations which may be solved to relate the outgoing amplitudes
$\beta$ and $\beta'$ to the incoming amplitudes $\alpha$ and $\alpha'$, \ie\ to
derive the $\cS$-matrix.
Using our previously derived results for $\phi$ and $\xhi$, we find that the above
equation reduces to
\begin{align}
0&=\begin{pmatrix} E & \gze\,S \\ \gze S^* & E \end{pmatrix}
\begin{pmatrix} \mu\,(\alpha+\beta') \\ (\beta+\alpha') \end{pmatrix}\nonumber\\
&\qquad
+\>\gon\begin{pmatrix}\big(\alpha\,e^{-i k\nd_z d} +\beta'\, e^{i k\nd_z d}
\big) \\ \mu\,\big(\beta\,e^{i k\nd_z d} + \alpha'\, e^{-i k\nd_z d}\big)
\end{pmatrix}\ .
\end{align}
This yields
\begin{align}
0&=\begin{pmatrix} \mu\, E + \gon\,e^{-i \theta\nd_z/2} & \gze\,S \\
\mu\,\gze\,S^* & E+\mu\, \gon\,e^{-i \theta\nd_z/2} \end{pmatrix}
\begin{pmatrix} \alpha \\ \alpha' \end{pmatrix}\nonumber\\
&\qquad\qquad +
\begin{pmatrix} \gze\,S & \mu\, E + \gon\,e^{i\theta\nd_z/2}  \\
E+\mu\, \gon\,e^{-i \theta\nd_z/2} & \mu\,\gze\,S^*  \end{pmatrix}
\begin{pmatrix} \beta \\ \beta' \end{pmatrix} \ ,
\end{align}
where $\theta\nd_z\equiv 2 k\nd_z d$.  The $\cS$-matrix is defined by
\begin{equation}
\begin{pmatrix} \beta \\ \beta' \end{pmatrix} = \stackrel{\cS-{\rm matrix}}
{\overbrace{\begin{pmatrix} t & r' \\ r & t' \end{pmatrix} }}
\begin{pmatrix} \alpha \\ \alpha' \end{pmatrix}\ .
\end{equation}
Solving for $\cS$, we obtain
\begin{equation}
\cS={-\begin{pmatrix} 2i\gze\,S^*\sin(\theta\nd_z/2) & \gon \\
\gon & 2i\gze\,S\sin(\theta\nd_z/2)  \end{pmatrix}
\over \gon\cos(\theta\nd_z) +2\, i\mu \,\sin(\theta\nd_z/2)
\left[\gamma_1^2\cos^2(\theta\nd_z/2) + \gamma_0^2\,|S|^2\right]^{1/2}}\ .
\end{equation}

\begin{figure}[!t]
\centering
\includegraphics[width=8cm]{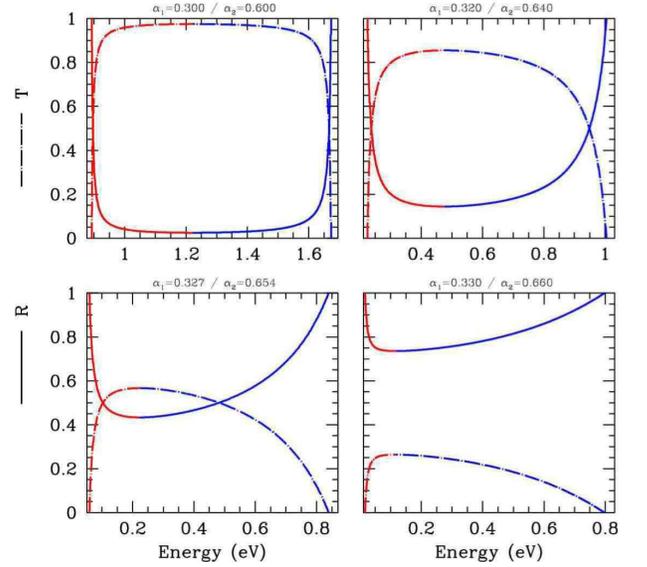}
\caption
{\label{trcoef} Reflection and transmission coefficients for $\bfk=\alpha_1\,\bfb_1
+\alpha_2\,\bfb_2$, for four sets of $(\alpha_1,\alpha_2)$, in the vicinity
of the Dirac point $(\frac{1}{3},\frac{2}{3})$.  Only positive energies are shown.}
\end{figure}

Thus, for all $\mu$ and $\mu'$, we have
\begin{equation}
R=|r|^2={\gamma_1^2\over \gamma_1^2 + 4\, \gamma_0^2\, |S|^2
\sin^2(\theta\nd_z/2)}\ ,
\end{equation}
and
\begin{equation}
T=|t|^2={4\, \gamma_0^2\, |S|^2\sin^2(\theta\nd_z/2)\over \gamma_1^2
 + 4\, \gamma_0^2\, |S|^2\sin^2(\theta\nd_z/2)}\ .
\end{equation}

As $\bfk$ approaches either zone corner $K$ or $K'$, the transmission goes to zero.
This is because the chains which extend through BHG are cut and shifted at the stacking fault.
Curiously, the transmission coefficient $T$ goes to unity when $\gon\to 0$.
Note also that along $K$-$H$ and $K'$-$H'$ we have $S=0$ and hence $R=1$, $T=0$.
At the band edges, we have
\begin{equation}
R(\theta\nd_z=0)=1 \quad,\quad
R(\theta\nd_z=\pi)={\gamma_1^2\over \gamma_1^2 + 4\, \gamma_0^2\,|S|^2} \ ,
\end{equation}
with $T=1-R$ for the transmission coefficients.

\subsection{Existence of Bound States}
To search for bound states, we take, for $j>0$,
\begin{align}
\psi\nd_{n=-2j}&=e^{\kappa n}\,\xhi \qquad & \psi\nd_{n=2j}&=\beta\,
e^{-\kappa n}\,\phi\\
\psi\nd_{n=-2j+1}&=e^{\kappa n}\,\phi \qquad & \psi\nd_{n=2j+1}&=\beta\,
e^{-\kappa n}\,\xhi\ ,
\end{align}
and solve for $\kappa$, $\beta$, and $E$.  At the plane $l=0$ we have
\begin{equation}
\psi\nd_0=\begin{pmatrix} \xhi\nd_1 \\ \beta\,\phi\nd_2\end{pmatrix}\ .
\end{equation}
The Schr{\"o}dinger equation for $l\ne 0$ then yields
\begin{align}
M\,\xhi + 2\gon\cosh(\kappa)\,\Sigma^+\phi&=0\\
M\,\phi + 2\gon\cosh(\kappa)\,\Sigma^-\xhi&=0\ .
\end{align}
This yields
\begin{equation}
E=-\mu\,\gon\cosh(\kappa)-\mu'\,\sqrt{\gamma_1^2\cosh^2(\kappa) + \gamma_0^2\,
|S|^2}\ ,
\end{equation}
where once again $\mu=\pm 1$ and $\mu'=\pm 1$.  Again we have
\begin{equation}
\phi=\begin{pmatrix} -\gze\,S \\ E \end{pmatrix} \qquad,\qquad
\xhi=\mu\begin{pmatrix} E \\ -\gze\,S^* \end{pmatrix}\ .
\end{equation}

\begin{figure}[!t]
\centering
\includegraphics[width=8cm]{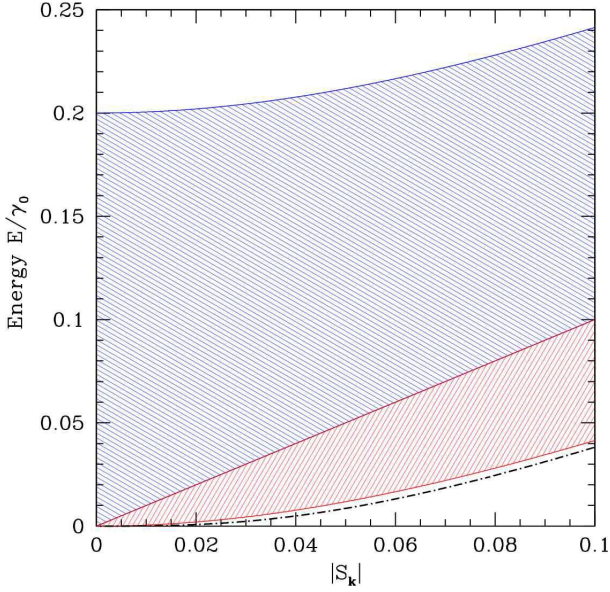}
\caption
{\label{bsa} Energy bands and bound state dispersion for $\gamma_1 = 0.1\,\gamma_0$
for small values of $|S|$.  The bulk bands $E^{\ssr(3,4)}_\bfk$ are depicted
by the red and blue hatched regions, respectively.  The bound state is the thick black
dot-dash curve.}
\end{figure}

At $l=0$ we again have
\begin{equation}
M\,\psi\nd_0 + \gon\,\Sigma^+\,\psi\nd_{-1} + \gon\,\Sigma^-\,\psi\nd_{+1}=0\ ,
\end{equation}
which here yields
\begin{equation}
\begin{pmatrix} E & \gze\,S \\ \gze\,S^* & E \end{pmatrix}
\begin{pmatrix} \mu \\ \beta \end{pmatrix}
+\gon\, e^{-\kappa}\begin{pmatrix} 1 \\ \mu\,\beta\end{pmatrix}=0\ .
\end{equation}
This yields two equations for $\beta$, which may be written as
\begin{equation}
\beta=-{\mu\, E + \gon e^{-\kappa}\over\gze\,S}=
-{\gze\,S^*\over \mu\, E +\gon e^{-\kappa}} \ .
\end{equation}
This fixes the energy at
\begin{equation}
E=-\mu\, \gon e^{-\kappa}\pm \gze\,|S|\ .
\end{equation}
Thus, we have a bound state at positive energy (and a corresponding one at negative
energy) for each real, positive value of $\kappa$, which solves one of the four
equations (for $\mu$, $\mu'=\pm 1$)
\begin{align}
-\mu\,\gon\,e^{-\kappa} + \mu'\,\gze\,|S| &= -\mu\,\gon\,\cosh(\kappa) \nonumber\\
&\qquad+ \mu'
\sqrt{\gamma_1^2\cosh^2(\kappa) + \gamma_0^2\,|S|^2}\ .
\label{BSE}
\end{align}
We assume $\gze>0$.  In the SWMC analysis \cite{SWMC}, one has
$\gon\approx -390\,$meV and $\gze\approx 3.2\,$eV.  The vertical hopping is
negative due to the sign of the overlap of $p\nd_z$ orbitals on consecutive layers.
In order to have a bound state solution, we must have $\mu\mu'={\rm sgn}(\gon)$, resulting in
\begin{equation}
\sqrt{\gamma_1^2\cosh^2(\kappa) + \gamma_0^2\,|S|^2} - \gze\,|S| =
\big|\gon\big|\,\sinh(\kappa)\ ,
\end{equation}
the solution of which is
\begin{equation}
\sinh(\kappa)={\big|\gon\big|\over 2\,\gze\,|S|}\equiv u\ .
\end{equation}

Thus, there are two bound states for all $\bfk$ in the Brillouin zone, one at positive
energy, corresponding to the choices $\mu=\mu'={\rm sgn}(\gon)$, and one at negative energy,
corresponding to the choices $\mu=\mu'=-{\rm sgn}(\gon)$.  Solving for $\kappa$, we have
\begin{equation}
e^{\pm\kappa}=\pm\, u+\sqrt{1+u^2}\ .
\end{equation}
The bound state energy may now be written as
\begin{align}
E\ns_\ssr{B}&=\big|\gon\big|\,\bigg(u+{1\over 2u}-\sqrt{1+u^2}\bigg)\vph\\
&={\gamma_0^3\, |S|^3\over \gamma_1^2} + \cO(u^{-5})\ ,\bvph
\end{align}
where the expansion in the second line is for large $u$, \ie\ $\gze\,|S|\ll \big|\gon\big|$.
Note that the bound state disperses as $|\bfk|^3$.  Recall for Bernal graphite that
the dispersion is linear in $|\bfk|$ in the vicinity of H and quadratic elsewhere along
the $K$-$H$ spine.  The length scale associated
with the bound state is $\kappa^{-1}$.   For $u\to\infty$, $\kappa^{-1} \sim 1/\ln(2u)$.

\begin{figure}[!t]
\centering
\includegraphics[width=8cm]{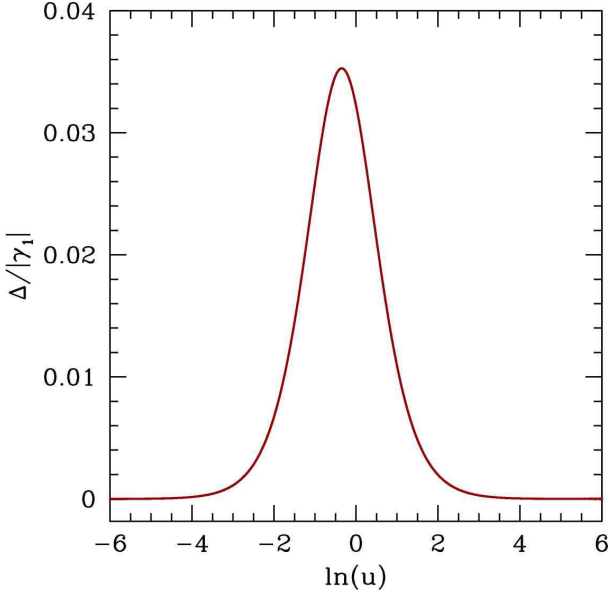}
\caption
{\label{delta} Binding energy of the bound state {\it versus\,} $\ln(u)$, where
$u=\big|\gamma_1/2\gamma_0 S\big|$.}
\end{figure}

Since the spectrum, including bound states, is particle-hole symmetric, we may
without loss of generality limit our attention to $E\ge 0$ states.  The continuum
bands, for fixed $\bfk$, range over energies
\begin{align}
-\big|\gon\big|+\sqrt{\gamma_1^2 + \gamma_0^2 \,|S|^2} \le & E^\ssr{(3)}_\bfk \le \gze\,|S| \\
\gze\,|S| \le & E^\ssr{(4)}_\bfk \le \big|\gon\big|+\sqrt{\gamma_1^2 + \gamma_0^2 \,|S|^2}\ .
\end{align}
The bound state we have analyzed lives just below the bottom of the $E^\ssr{(3)}_\bfk$
band.  The binding energy is $\Delta=E^{\ssr(3)}_\ssr{min}-E\ns_\ssr{B}$, and is given by
\begin{equation}
{\Delta(u)\over |\gon|}={1\over 2u}\Big(\sqrt{1+4u^2}-1\Big)+\sqrt{1+u^2}-1-u\ .
\end{equation}
In figs. \ref{bsa} we plot the bound state spectrum for the case $|\gon|/\gze=0.1$
for small values of $|S|$, \ie\ close to the zone corners, where $u$ is large.
At the zone center, $|S|=3$ is maximized and $u$ achieves its minimum value;
for reference, $u\ns_\ssr{SWMC}=0.02057$.
The binding energy vanishes in both the $u\to 0$ and $u\to\infty$ limits, as shown
in fig. \ref{delta}.  The maximum of $\Delta$ occurs for $u=1$, where
$\Delta/|\gon|=0.03225$, corresponding to a binding energy of approximately
$13\,$meV.  In the appendix, we compute the bound state spectrum for the full
SWMC model.

\section{Finite B Case}
To obtain the Landau levels, we expand about the Dirac points, following the
method described in section \ref{weak}.  We have $S\nd_{\bfK+\bfpi/\hbar}=-\eps\,b$,
with $\eps$ given in eqn. \ref{epse}.  At $B=10\,\rmT$ one has $\eps=0.0371$.  With $\gon=0.39\,$eV and $\gze=3.16\,$eV,
we have $r=0.123$ and $\eps/r=\sqrt{B/B\ns_1}$ where $B\ns_1=110.8\,\rmT$.
For physical fields, then, we have $\eps \ltwid r$.  Note that one can also write
\begin{equation}
\eps\,\gze={\sqrt{2}\,\hbar\,\vF\over\ell_B}\ ,
\end{equation}
where $\vF=\sqrt{3}\,\gze\,a/2\hbar$ is the Fermi velocity ($a=2.46\,$\AA\ is the
lattice spacing in the hexagonal planes) and $\ellb=\hbar c/eB$ is the magnetic length.

\subsection{Bernal Stacking and Landau Levels}
We define the operator-valued matrix
\begin{equation}
\Mhat=\begin{pmatrix} E & \eps\,\gze\,b \\ \eps\,\gze\,b\yd & E \end{pmatrix}\ .
\end{equation}
For perfect Bernal stacking, we have
\begin{align}
\Mhat\,\psi\nd_{2j} + \gon\,\Sigma^+\big(\psi\nd_{2j-1}+\psi\nd_{2j+1}\big)&=0\\
\Mhat\,\psi\nd_{2j+1} + \gon\,\Sigma^-\big(\psi\nd_{2j}+\psi\nd_{2j+2}\big)&=0\ .
\end{align}
We now write the wavefunction in terms of right and left moving components:
\begin{align}
\psi\nd_{2j}&=\big( I\,e^{iq j} + O'\,e^{-iq j}\big)\begin{pmatrix} \alpha\,\ket{n} \\
\beta\,\ket{n+1}\end{pmatrix}\\
\psi\nd_{2j+1}&=\big( I\,e^{iq (j+{1\over 2})} + O'\,e^{-iq (j+{1\over 2})}\big)
\begin{pmatrix} x\,\ket{n-1} \\ y\,\ket{n}\end{pmatrix}\ ,
\label{abxy}
\end{align}
where we assume $n>0$.  We therefore have
\begin{align}
M\nd_n \begin{pmatrix} x \\ y \end{pmatrix}
+2\gon\cos(q/2)\,\Sigma^-\begin{pmatrix} \alpha \\ \beta \end{pmatrix}&=0\\
M\nd_{n+1} \begin{pmatrix} \alpha \\ \beta \end{pmatrix}
+2\gon\cos(\ql/2)\,\Sigma^+\begin{pmatrix} x \\ y \end{pmatrix}&=0
\end{align}
where
\begin{equation}
M\nd_n\equiv \begin{pmatrix} E & \sqrt{n}\>\eps\,\gze\\
\sqrt{n}\>\eps\,\gze & E \end{pmatrix}\ .
\end{equation}
This leads to
\begin{align}
P\nd_n(E)&={\rm det}\,\Big[M\nd_{n+1}-4\gamma_1^2\cos^2(q/2)\,\Sigma^+\,
M^{-1}_n\,\Sigma^-\Big]\nonumber\\
&=E^2-(n+1)\,\eps^2\,\gamma_0^2 - {4 \gamma_1^2E^2\cos^2(q/2)\over
E^2-n\,\eps^2\,\gamma_0^2}\ .
\end{align}
Setting $P\nd_n(E)=0$ yields the spectrum $E=E\nd_n(q)$ of Bernal
hexagonal graphite:
\begin{align}
&{E_{n,\pm}^2(q)\over \gamma_0^2}=(n+\half)\,\eps^2 + 2r^2\cos^2(q/2)\\
&\quad\pm\sqrt{4 r^4\cos^4(q/2) + (4n+2)\,\eps^2\,r^2\cos^2(q/2)
+\frac{1}{4}\,\eps^4}\ ,\nonumber
\end{align}
where $r=\gon/\gze$.  Expanding for small $\eps/r$, we have
\begin{align}
E\nd_{n,-}&\in\bigg[\sqrt{n(n+1)}\ {\eps^2\over 2r}\ ,
\ \eps\sqrt{n}\bigg]\\
E\nd_{n,+}&\in \bigg[\sqrt{n+1} \ \eps\ ,\ 2r+\big(n+\half\big)\,
{\eps^2\over 2r} + \ldots\bigg]
\end{align}

\subsection{Zero Modes}
The case $n=0$ must be considered separately.  Consider the wavefunction
\begin{equation}
\psi\nd_{2j}=\begin{pmatrix} 0 \\ \beta\,\ket{0} \end{pmatrix}\,\delta\nd_{j,J}
\quad,\quad
\psi_{2j+1}=\begin{pmatrix} 0 \\ 0 \end{pmatrix}\ .
\end{equation}
This is an $E=0$ eigenstate for any $J$.  It describes a state localized on a single
plane.

\begin{figure}[!t]
\centering
\includegraphics[width=8cm]{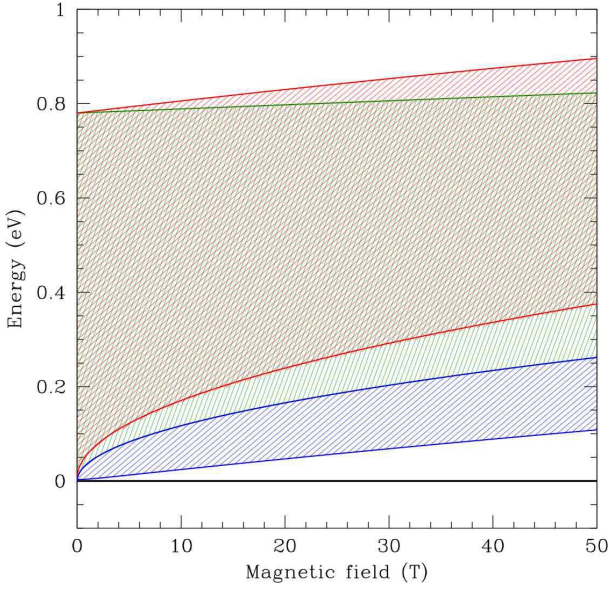}
\caption
{\label{GLLs} Landau levels in graphite.  Subbands $E_{n=1,+}$ (red),
$E_{n=1,-}$ (blue), and $E_{n=0,+}$ (green) are shown.  The zero modes
are shown in black.}
\end{figure}

We can find more solutions by writing
\begin{align}
\psi\nd_{2j}&=\begin{pmatrix} \alpha\,\ket{0} \\ \beta\,\ket{1} \end{pmatrix} e^{iqj} \\
\psi\nd_{2j+1}&=\begin{pmatrix} 0 \\ y\,\ket{0} \end{pmatrix}e^{iq (j+{1\over 2})}  \ .
\end{align}
The Schr{\"o}dinger equation then requires
\begin{equation}
\begin{pmatrix} E & \eps\,\gze \\ \eps\,\gze & E \end{pmatrix}
\begin{pmatrix} \alpha \\ \beta \end{pmatrix} + 2\gon\cos(q/2)
\begin{pmatrix} y \\ 0 \end{pmatrix}=0
\end{equation}
on even planes and
\begin{equation}
E\begin{pmatrix} 0 \\ y \end{pmatrix} + 2\gon\cos(q/2)
\begin{pmatrix} 0 \\ \alpha \end{pmatrix}=0
\end{equation}
on odd planes.  Thus, we have three equations for the remaining three eigenvalues:
\begin{align}
0&=E\,\alpha + \eps\,\gze\,\beta + 2\gon\cos(q/2)\,y \\
0&=E\,\beta + \eps\,\gze\,\alpha \\
0&=E\, y + 2\gon\cos(q/2)\,\alpha\ .
\end{align}
We immediately see that $E=0$ is an eigenvalue, with eigenvector
\begin{equation}
\begin{pmatrix} \alpha \\ \beta \\ y \end{pmatrix} = \begin{pmatrix} 0 \\ -2\gon\cos(q/2) \\
\eps\,\gze \end{pmatrix}\ .
\end{equation}
If we Fourier transform this solution, multiplying by $e^{-iq(J+{1\over 2})}$ and summing
over $q$, we find a purely localized state, with
\begin{align}
\psi\nd_{2J}&=-\gon\begin{pmatrix} 0 \\ \ket{1} \end{pmatrix} \\
\psi\nd_{2J+1}&=\eps\,\gze\begin{pmatrix} 0 \\ \ket{0} \end{pmatrix} \\
\psi\nd_{2J+2}&=-\gon\begin{pmatrix} 0 \\ \ket{1} \end{pmatrix} \ ,
\end{align}
with all other $\psi\nd_n=0$.  This zero mode is localized on three layers.
The remaining two solutions are easily found to be
\begin{equation}
\begin{pmatrix} \alpha \\ \beta \\ y \end{pmatrix} =
\begin{pmatrix} E \\ -\eps\,\gze \\ -2\gon\cos(q/2) \end{pmatrix}\ ,
\end{equation}
with $E=E\nd_{0,\pm}\equiv\pm\sqrt{\eps^2 \gamma_0^2 + 4 \gamma_1^2\cos^2(q/2)}$.
These solutions are wave-like and disperse with $q$.

\begin{figure}[!t]
\centering
\includegraphics[width=8cm]{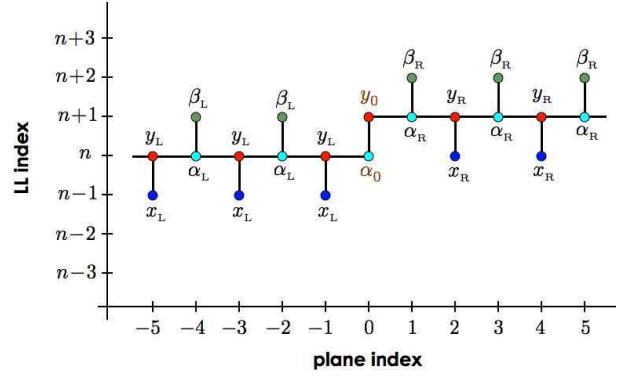}
\caption
{\label{LL_step} Landau level indices for scattering at a stacking fault in Bernal
graphite.}
\end{figure}

\subsection{Stacking Fault}
For the system with a single step stacking fault, the situation is as depicted in
fig. \ref{LL_step}.  We then swap the notation for even and odd planes on the right
half of the system (layer indices $l>0$) with respect to eqn. \ref{abxy}, and introduce
wavevectors $\ql$ and $\qr$ for the left and right half-systems.
We must then match the energies on left and right sides of the fault:
\begin{equation}
E\nd_n(\ql)=E\nd_{n+1}(\qr)\ .
\end{equation}

To identify the bound states, we write the wavefunction for $l>0$ as
\begin{equation}
\psi\nd_{2j-1}=\begin{pmatrix} \alpha\nd_j\,\ket{n+1} \\ \beta\nd_j\,\ket{n+2}
\end{pmatrix} \quad,\quad
\psi\nd_{2j}=\begin{pmatrix} x\nd_j\,\ket{n} \\  y\nd_j\ket{n+1} \end{pmatrix}
\end{equation}
and for $l<0$ as
\begin{equation}
\psi\nd_{-(2j-1)}=\begin{pmatrix} \abar\nd_j\,\ket{n-1} \\ \bbar\nd_j\,\ket{n}
\end{pmatrix} \quad,\quad
\psi\nd_{-2j}=\begin{pmatrix} \xbar\nd_j\,\ket{n} \\  \ybar\nd_j\ket{n+1} \end{pmatrix}\ .
\end{equation}
At $l=0$ we write
\begin{equation}
\psi\nd_0=\begin{pmatrix} \xbar\nd_0\,\ket{n} \\ y\nd_0\ket{n+1}\end{pmatrix}\ .
\end{equation}
The Schr{\"o}dinger equation, evaluated for both even and odd planes with $l>0$
and $l<0$ now gives eight relations among the eight sets of coefficients
$\big\{\alpha\nd_j,\beta\nd_j,x\nd_j,y\nd_j,\abar\nd_j,\bbar\nd_j,\xbar\nd_j,\ybar\nd_j\big\}$,
expressible as
\begin{equation}
\begin{pmatrix} E & \sqrt{n+1}\ \eps\,\gze \\  \sqrt{n+1}\ \eps\,\gze & E\end{pmatrix}
\begin{pmatrix} x\ns_j \\ y\nd_j \end{pmatrix}
+ \gon\begin{pmatrix} 0 \\ \alpha\nd_j+\alpha\nd_{j+1} \end{pmatrix}=0
\end{equation}
and
\begin{equation}
\begin{pmatrix} E & \sqrt{n+2}\ \eps\,\gze \\  \sqrt{n+2}\ \eps\,\gze & E\end{pmatrix}
\begin{pmatrix} \alpha\ns_j \\ \beta\nd_j \end{pmatrix}
+ \gon\begin{pmatrix} y\ns_{j-1}+y\ns_j \\ 0 \end{pmatrix}=0
\end{equation}
and
\begin{equation}
\begin{pmatrix} E & \sqrt{n+1}\ \eps\,\gze \\  \sqrt{n+1}\ \eps\,\gze & E\end{pmatrix}
\begin{pmatrix} \xbar\ns_j \\ \ybar\nd_j \end{pmatrix}
+ \gon\begin{pmatrix} \bbar\nd_j+\bbar\nd_{j+1} \\ 0 \end{pmatrix}=0
\end{equation}
and
\begin{equation}
\begin{pmatrix} E & \sqrt{n}\ \eps\,\gze \\  \sqrt{n}\ \eps\,\gze & E\end{pmatrix}
\begin{pmatrix} \abar\ns_j \\ \bbar\nd_j \end{pmatrix}
+ \gon\begin{pmatrix}  0 \\ \xbar\nd_{j-1}+\xbar\nd_j  \end{pmatrix}=0\ .
\end{equation}

We can use these equations to eliminate the four sets of coefficients
$\{\beta\nd_j,x\nd_j,\abar\nd_j,\ybar\nd_j\}$:
\begin{align}
\beta\nd_j&=-\sqrt{n+2}\ \eps\,\gze\,E^{-1}\,\alpha\nd_j \\
\ybar\nd_j&=-\sqrt{n+1}\ \eps\,\gze\,E^{-1}\,\xbar\nd_j \\
x\nd_j&=-\sqrt{n+1}\ \eps\,\gze\,E^{-1}\,y\nd_j \\
\abar\nd_j&=-\sqrt{n}\ \eps\,\gze\,E^{-1}\,\bbar\nd_j\ .
\end{align}
We then obtain
\begin{align}
0&=R\nd_{n+1}(E)\,y\nd_j + \alpha\nd_j+\alpha\nd_{j+1}\\
0&=R\nd_{n+2}(E)\,\alpha\nd_j + y\nd_{j-1}+y\nd_j\\
0&=R\nd_{n+1}(E)\,\xbar\nd_j + \bbar\nd_j+\bbar\nd_{j+1}\\
0&=R\nd_{n}(E)\,\bbar\nd_j + \xbar\nd_{j-1}+\xbar\nd_j\ ,
\end{align}
where
\begin{equation}
R\nd_n(E)\equiv {E^2-E_n^2\over E\,\gon}\ ,
\end{equation}
with $E^2_n\equiv n\,\eps^2\,\gamma_0^2$.
We then have
\begin{equation}
\begin{pmatrix} \alpha\nd_{j+1} \\ y\nd_j \end{pmatrix} = \big(K_{n+1}\big)^j
\begin{pmatrix} \alpha\nd_1 \\ y\nd_0 \end{pmatrix}
\end{equation}
and
\begin{equation}
\begin{pmatrix} \bbar\nd_{j+1} \\ \xbar\nd_j \end{pmatrix} = \big(K_n\big)^j
\begin{pmatrix} \bbar\nd_1 \\ \xbar\nd_0 \end{pmatrix}\ ,
\label{soln}
\end{equation}
where
\begin{equation}
K\nd_n(E)=\begin{pmatrix} R\nd_n(E)\,R\nd_{n+1}(E) -1 & R\nd_n(E) \\
-R\nd_{n+1}(E) & -1 \end{pmatrix}\ .
\end{equation}

Note that ${\rm det}\,K\nd_n(E)=1$, and that the characteristic polynomial
${\rm det}\,(\lambda-K\nd_n)$ is real for real $\lambda$.  It is easy to see
that the eigenvalues of $K\nd_n(E)$ form a complex conjugate pair $e^{\pm i\theta}$
if the energy $E$ satisfies the condition  the condition
$\big|{\rm Tr}\,K\nd_n(E)\big| \le 2$, or
\begin{equation}
0\le R\nd_n(E)\,R\nd_{n+1}(E) \le 4\ .
\end{equation}
This is the condition that $E$ lies within one of four energy bands.
The roots of $R\nd_n(E)\,R\nd_{n+1}(E)=0$ lie at $E^2=E_n^2$ and
$E^2=E_{n+1}^2$, while the roots of $R\nd_n(E)\,R\nd_{n+1}(E)=4$
lie at $E^2=E_{n,-}^2$ and  $E^2=E_{n+1,+}^2$, where
\begin{align}
E_{n,\pm}^2&=(n+\half)\,\eps^2\, \gamma_0^2 + 2\,\gamma_1^2\\
&\qquad\qquad\pm
\sqrt{4\, \gamma_1^4 + (4n+2)\, \eps^2\,\gamma_0^2\,\gamma_1^2 + \frac{1}{4} \,\eps^4\,\gamma_0^4}\ .\nonumber
\end{align}
The bands are then given by
\begin{equation}
E_{n,-}^2 \le E^2 \le E_n^2 \qquad,\qquad
E_{n+1}^2 \le E^2 \le E_{n,+}^2\ .
\end{equation}
In the limit $\sigma\equiv \eps^2\,\gamma_0^2/\gamma_1^2\ll 1$, we can expand
and write
\begin{align}
E_{n,-}^2&\simeq n(n+1)\,{\eps^4\,\gamma_0^4\over 4\,\gamma_1^2}\\
E_{n,+}&\simeq 4\,\gamma_1^2 + (2n+1)\,\eps^2\,\gamma_0^2\ .
\end{align}

At plane $n=0$ the Schr{\"o}dinger equation yields
\begin{equation}
\begin{pmatrix} \gon & E \\ 0 & \sqrt{n+1}\ \eps\,\gze\end{pmatrix}
\begin{pmatrix} \bbar\nd_1 \\ \xbar\nd_0\end{pmatrix} +
\begin{pmatrix} 0 & \sqrt{n+1}\ \eps\,\gze \\ \gon & E \\ \end{pmatrix}
\begin{pmatrix} \alpha\nd_1 \\ y\nd_0\end{pmatrix} =0\ .
\label{central}
\end{equation}

\subsection{Scattering Matrix}
If both $\big|{\rm Tr}\,K\nd_n(E)\big|\le 2$ and $\big|{\rm Tr}\,K\nd_{n+1}(E)\big|\le 2$,
then we can write
\begin{equation}
\begin{pmatrix} \bbar\nd_1 \\ \xbar\nd_0 \end{pmatrix} =  I\,\rmPsi_-^{(n)} +
O'\,\rmPsi_+^{(n)}
\label{left}
\end{equation}
and
\begin{equation}
\begin{pmatrix} \alpha\nd_1 \\ y\nd_0 \end{pmatrix} =  I'\,\rmPsi_-^{(n+1)} +
O\,\rmPsi_+^{(n+1)},
\label{right}
\end{equation}
where
\begin{equation}
K\nd_n(E)\,\rmPsi_\pm^{(n)}=e^{\pm i\theta\nd_n}\,\rmPsi_\pm^{(n)}\ .
\end{equation}
Then we have
\begin{align}
\begin{pmatrix} \bbar\nd_{j+1} \\ \xbar\nd_j \end{pmatrix} &=
I\,e^{-ij\theta\nd_n}\,\rmPsi_-^{(n)} + O'\,e^{ij\theta\nd_n}\,\rmPsi_+^{(n)} \\
\begin{pmatrix} \alpha\nd_{j+1} \\ y\nd_j \end{pmatrix} &=
I'\,e^{-ij\theta\nd_{n+1}}\,\rmPsi_-^{(n+1)} + O\,e^{ij\theta\nd_{n+1}}\,\rmPsi_+^{(n+1)}\ .
\end{align}
The $\cS$-matrix, which relates incoming to outgoing flux amplitudes, is then obtained from eqns.
\ref{central}, \ref{left}, and \ref{right}, upon replacing $I\to v^{1/2}_\ssr{L}\,\cI$, $O \to \cO' v^{1/2}_\ssr{L}$,
$I'\to v^{1/2}_\ssr{R}\,\cI$, and $O \to \cO v^{1/2}_\ssr{R}$, where $v\ns_\ssr{L}=\pz E_n(q\ns_\ssr{L})/
\pz q\ns_\ssr{L}$ and $v\ns_\ssr{R}=\pz E_{n+1}(q\ns_\ssr{R})/\pz q\ns_\ssr{R}$

\begin{figure}[!t]
\centering
\includegraphics[width=8cm]{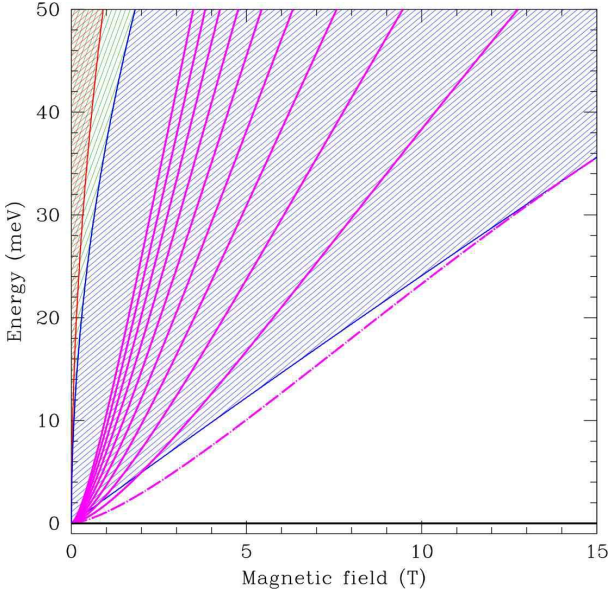}
\caption
{\label{bsb} Bulk energy bands (shaded and hatched regions) and bound states
(magenta curves) {\it versus\/} magnetic field for $\gamma_0=3.16\,$eV and
$\gamma_1=-0.39\,$eV (tight binding; nearest neighbor hopping only).
The lowest energy bound state merges into the band
continuum at $B\approx 15\,$T.  The other bound states remain sharp over the
energy range shown and do not mix with the lowest bulk band. }
\end{figure}

\subsection{Bound States}
If a state is evanescent on both sides of the stacking fault, we must have that
both $\big|{\rm Tr}\,K\nd_n(E)\big| > 2$ and $\big|{\rm Tr}\,K\nd_{n+1}(E)\big| > 2$.
The eigenvalues of $K\nd_n(E)$ are given by
\begin{equation}
\Lambda\nd_{n,\pm}=\half\,\tau\nd_n \pm\half\sqrt{\tau_n^2-4}\ ,
\end{equation}
where
\begin{equation}
\tau\nd_n(E)\equiv {\rm Tr}\,K\nd_n(E)=R\nd_n(E)\,R\nd_{n+1}(E)-2\ .
\end{equation}
In order that the solution in eqn. \ref{soln} not blow up for $n\to\pm\infty$, we
must require that $\begin{pmatrix} \alpha\nd_1 \\ y\nd_0\end{pmatrix}$
and $\begin{pmatrix} \bbar\nd_1 \\ \xbar\nd_0\end{pmatrix}$ have no weight in the
$|\Lambda|>1$ eigenspaces for $K\nd_n(E)$ and $K\nd_{n+1}(E)$, respectively.  This
means
\begin{align}
R\nd_{n+2}(E)\,\alpha\nd_1 + y\nd_0&=-\Lambda^<_{n+1}\,y\nd_0\vph\\
R\nd_{n+1}(E)\,\bbar\nd_1 + \xbar\nd_0&=-\Lambda^<_{n}\,\xbar\nd_0\ ,
\end{align}
where $|\Lambda^<|<1$.
When we combine these equations with those in eqn. \ref{central}, we obtain
\begin{equation}
\cM\begin{pmatrix} \alpha\nd_1 \\ y\nd_0 \\ \bbar\nd_1 \\ \xbar\nd_0
\end{pmatrix}=0\ .
\end{equation}
where
\begin{equation}
\cM=\begin{pmatrix}
R\nd_{n+2} & 1+\Lambda^<_{n+1} & 0 & 0 \\
\gon & E & 0 & \sqrt{n+1}\ \eps\,\gze \\
0 & 0 & R\nd_{n+1} & 1+\Lambda^<_{n} \\
0 & \sqrt{n+1}\ \eps\,\gze  & \gon & E \end{pmatrix}\ .
\end{equation}
A solution requires $D(E)={\rm det}\,\cM(E)=0$.  We have
\begin{align}
D(E)&=\big[E\,R\nd_{n+2}-\gon(1+\Lambda^<_{n+1})\big]\,
\big[E\,R\nd_{n+1}-\gon(1+\Lambda^<_{n})\big]\nonumber\\
&\qquad-(n+1)\,\eps^2\,\gamma_0^2\,R\nd_{n+1}\,R\nd_{n+2}\ .
\end{align}

Let us look for a bound state with energy $E$ which is parametrically (in $\sigma$)
smaller than both $\gon$ and $\eps\,\gze$.  Then $R\nd_n(E)\simeq -E_n^2/E\,\gon$, from
which we obtain $\Lambda^<_n \simeq \big(R\nd_n\,R\nd_{n+1}\big)^{-1}$.  Then find
\begin{equation}
D(E)\approx \gamma_1^2-(n+1)^2(n+2)\,{\eps^6\,\gamma_0^6\over \gamma_1^2\,E^2}\ .
\end{equation}
Setting $D(E)=0$ yields the bound state energy,
\begin{equation}
E^2=(n+1)^2(n+2)\,{\eps^6\,\gamma_0^6\over \gamma_1^4}\ .
\end{equation}
Thus, the bound state energy is proportional to $|B|^{3/2}$.
In fig. \ref{bsb}, we plot the lowest ten bound state energies {\it versus\/} magnetic field.

\section{Surface Spectroscopy of Buried Stacking Faults}
Our previous results for the transmission through a stacking defect
suggest that these defects are very effective in decoupling graphene
stacks. We analyze now the density of states at a graphite surface
in the presence of a stacking defect a few layers below the surface.
The stacking sequence is (AB)$\ns_{N/2}$CBCB$\cdots$. The number of layers between
the surface and the defect is $N$.

\begin{figure}
\begin{center}
\includegraphics*[width=8cm,angle=0]{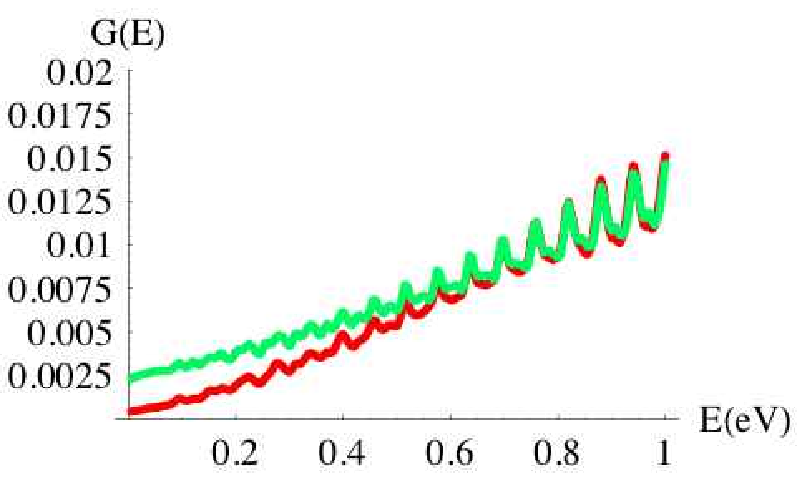}
\includegraphics*[width=8cm,angle=0]{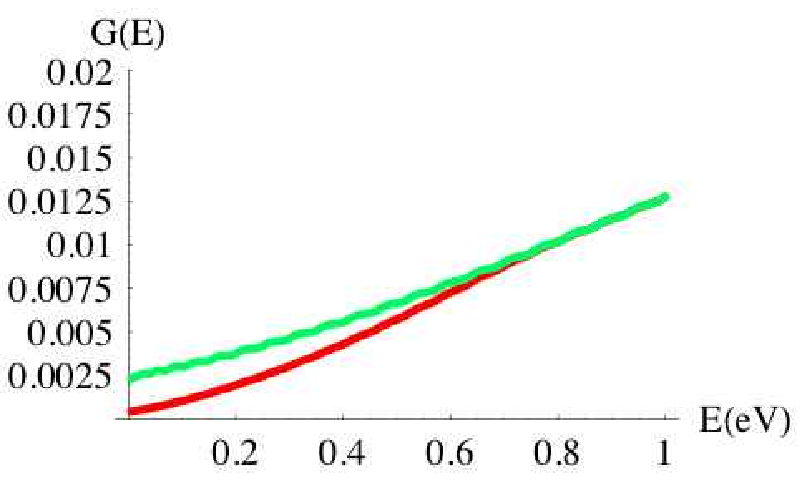}
 \end{center}
\caption{(Color online). Left: Density of states for the two
inequivalent sites of a graphite surface with a stacking defect 20
layers below the surface Triangles (red) give the density of states
at the site with a nearest neighbor in the layer below. Squares
(green) give the density of states at the site without nearest
neighbors in the layer below. Right: as in the left panel, with a
defect 100 layers below the surface. }
  \label{green0}
\end{figure}

The system can be separated into a perfect semi-infinite graphite
sample coupled to the defect layer, and $N$ layers between the
defect and the surface. We will only include the parameters
$\gze$ and $\gon$.  The semi-infinite portion can be integrated out.
The site of the defect layer connected to it acquires a self energy:
\begin{equation}
\Sigma_0 ( \omega ) = {2 \gamma_1^2\over \left( \omega -
{|\gze\,S|^2\over \omega} \right) - \sqrt{ \left( \omega -
{|\gze\,S|^2\over \omega} \right)^2 - 4 \gamma_1^2}}
\label{semiinfinite}
\end{equation}
We now integrate out this site, leading to the self energy:
\begin{equation}
\Sigma_1 ( \omega ) ={|\gze\,S|^2\over\omega - \Sigma_0 (
\omega ) }
\end{equation}
The procedure can be iterated leading to new self energies for sites
$2$, $3\ldots N$, resulting in the hierarchy
\begin{equation}
\Sigma_{i+1} ( \omega ) = {|\gze\,S|^2\over\omega} +
{\gamma_1^2\over \omega - \Sigma_i ( \omega )}
\end{equation}
The Green's function at the two inequivalent sites of the surface
layer ($N$) are:
\begin{equation}
G_{\rm u}^{N} ( \omega ) = {1\over \omega -
{|\gze\,S|^2\over \omega} - {\gamma_1^2\over \omega - \Sigma\nd_{N-1} (
\omega )}}
\end{equation}
and
\begin{equation}
G_{\rm v}^{N} ( \omega ) = {1 \over \omega -
{|\gze\,S|^2\over \omega - {\gamma_1^2\over \omega - \Sigma\nd_{N-1} (
\omega )}}} \label{green_s}
\end{equation}

We show in fig. \ref{green0} the surface
density of states when such a defect lies twenty and hundred layers
below the surface, obtained by integrating  the
imaginary part of the Green's functions in eq.~\ref{green_s}
over the in-plane component $\bfk_\parallel$ of the wavevector.

\begin{figure}
\begin{center}
\includegraphics*[width=6cm,angle=0]{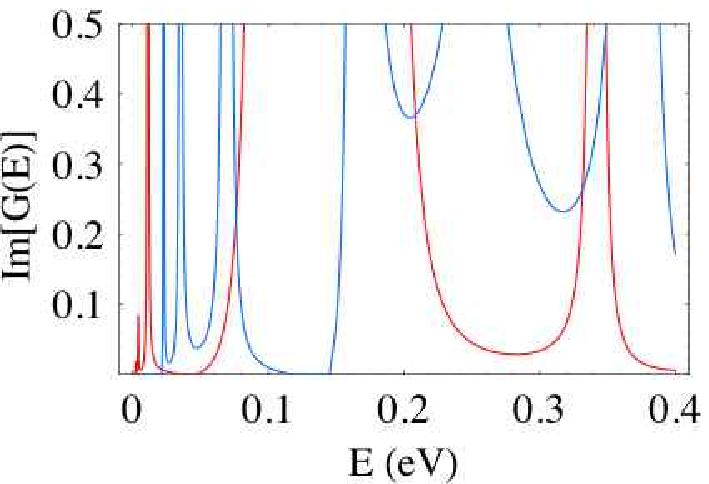}
\includegraphics*[width=6cm,angle=0]{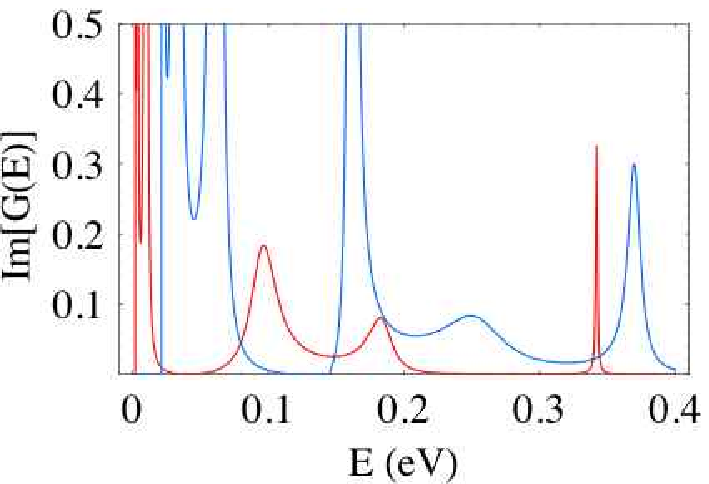}
 \end{center}
\caption{(Color online). Surface density of states for a
semiinfinite stack with a defect  ten layers below the surface. The
Landau level index is $n=2$, and the fields studied are B = 1 T
(red) and B = 10 T (blue).
  Left: Sublattice with a nearest neighbor in the contiguous layer.
  Right: Sublattice without a neighbor in the contiguous layer.}
  \label{green}
\end{figure}

The density of states show a number of resonances, which are
smoothed out when the number of layers between the defect and the
surface is large. For $N \gg 1$, we recover the analytical results
in ref. \cite{GNP06}. These results are consistent with the analysis in
the previous sections, which show that the transmission through the
defect is strongly suppressed. The layers between the defect and the
surface become effectively decoupled from the bulk of the system.

\begin{figure}
\begin{center}
\includegraphics*[width=6cm,angle=0]{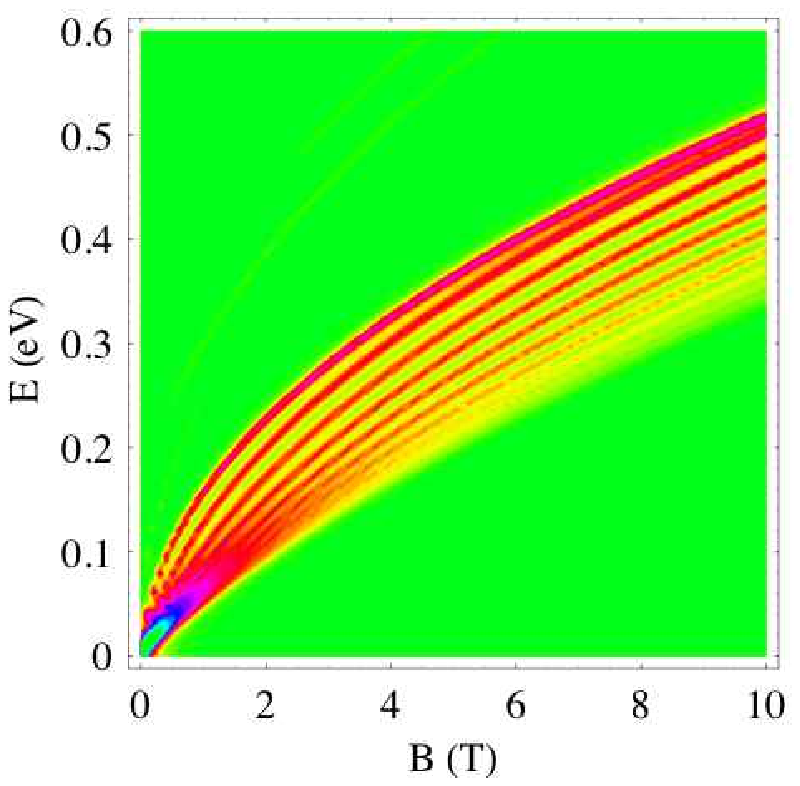}
\includegraphics*[width=6cm,angle=0]{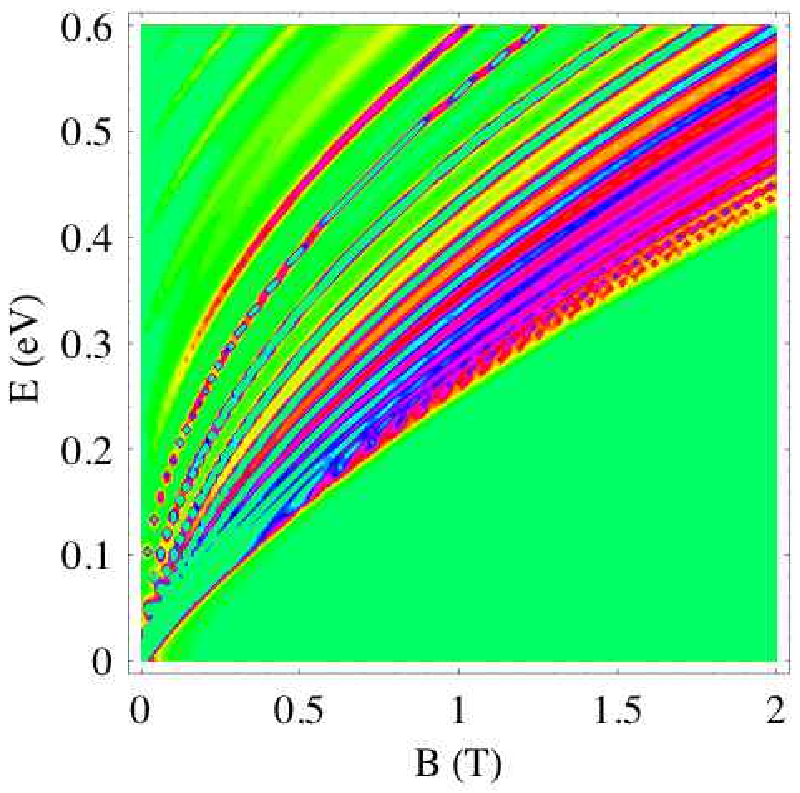}
 \end{center}
\caption{Surface density of states at the sublattice without a
nearest neighbor in the next layer. The system has a stacking
fault of the type described in the text ten layers from the surface.
Top: $n=2$. Bottom: $n=10$.} \label{green_3D}
\end{figure}

The previous analysis can be extended to the study of Landau levels
in a magnetic field. As discussed earlier, the hoppings within the
layers depend now on the Landau level index $n$, instead of on $\bfk\ns_\parallel$.
The $n$ dependence of the hoppings in he two
layers within the unit cell is different. Because of this, the self
energy obtained by integrating out the perfect semiinfinite region
leads to a more complicated expression than those in
eq.~\ref{semiinfinite}. Within the region between the defect and the
surface the successive self energies have a twofold periodicity:
\begin{align}
\Sigma_i ( \omega ) &={n\,v_\ssr{F}^2\, \ell_B^{-2}\over \omega} +
{\gamma_1^2\over \omega - \Sigma_{i-1} ( \omega )} \\
\Sigma_{i+1} ( \omega ) &= {(n-1)\,v_\ssr{F}^2\, \ell_B^{-2} \over \omega} +
{\gamma_1^2\over \omega - \Sigma_{i} ( \omega )}
\end{align}
The resulting densities of states for Landau level index $n=2$ and
two magnetic fields, $B=1\,$T and $B=10\,$T, are shown in
fig.~\ref{green}.

We show finally in fig. [\ref{green_3D}] the dependence of the peaks
in the surface density of states on the magnetic field. As before,
there is a stacking defects ten layers below the surface. In
agreement with experiments \cite{MAT05,NIM06,LI07}, there are
peaks which scale as $\sqrt{B}$ and peaks which scale as $B$.

\section{Discussion}
We have analyzed the appearance of two dimensional features in bulk
graphite.  We show that deviations from the Bernal stacking order are
very effective in inducing two dimensional behavior.  An ordered
array of graphene layers with the rhombohedral stacking order leads
to isolated Landau levels, and to quantized quantum Hall
plateaus at moderate magnetic fields in doped systems.  We found
that the gap between Landau level subbands of indices $n$ and $n+1$
opens at a field $B_{\rmc,n}$ with $B_{\rmc,n=0}\equiv B\ns_0=0.123\,\rmT$
and $B_{\rmc,n}\sim 4n\,B_0$ for large $n$.  By contrast, in Bernal
graphite, the first gap is predicted to open at fields on the order of $10\,$T
\cite{BHRA07}, and the second gap opens only at enormous field, on the order
of 1000 T.

We have also considered the simplest stacking defect in Bernal
graphite, which has locally a rhombohedral arrangement.  These
defects are expected to be common in many graphite samples, and
concentrations up to 10\% have been reported~\cite{BCP88}. These
defects are very effective in decoupling the electronic states on
either side.  They also give rise to a two dimensional band
of electronic states, localized in the vicinity of the defect.
Within a nearest neighbor tight binding model for the $\pi$ band
of graphite, with in-plane hopping $\gze=3.16\,$eV and interplane
hopping $\gon=0.39\,$eV, we found a maximum binding energy of
approximately 13 meV, for states rather close to the corners in the
basal Brillouin zone.  When the full SWMC model is taken into account
\cite{SWMC}, we obtain a maximum binding energy of almost 40 meV
for electron states and 20 meV for hole states; the binding energy is
significant only along the zone faces.

\begin{figure}[!t]
\begin{center}
\includegraphics*[width=7cm,angle=0]{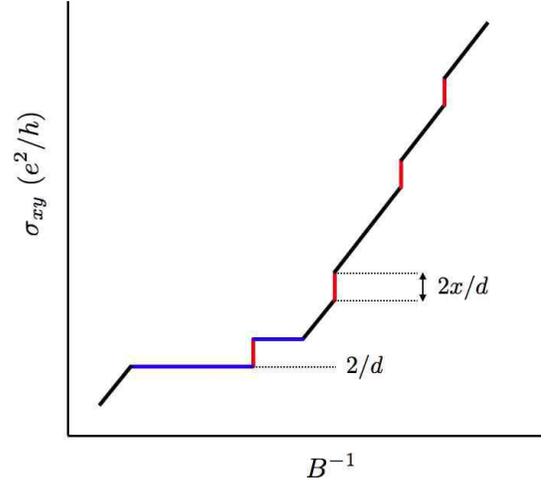}
 \end{center}
\caption{Sketch of the expected behavior of the Hall conductivity in
the quantum limit, for a lightly doped system.  The leftmost plateau is
a bulk effect, related to the Landau levels of Bernal graphite~\cite{BHRA07}. The size of the
other jumps depends on the concentration $x$ of stacking defects.
The continuum bands of Landau levels lead to a monotonically varying conductivity.}
\label{sketch_QHE}
\end{figure}

What are the implications of our work for magnetotransport in graphite with stacking faults?
To describe the physics, it is helpful to keep in mind the bound state Landau level
structure of fig. \ref{bsb}.  First, suppose the graphite is undoped.  In this case, the Fermi
level remains pinned within the central $n=0$ Landau levels.   With only nearest neighbor
hoppings, there are two flat bands (\ie\ which do not disperse as a function of $k\ns_z$)
at $E=0$ associated with each zone corner in the basal Brillouin zone.  Taking into account
the weak second-neighbor plane hoppings $\gtw$ and $\gfi$, these bands disperse and
acquire a width of about 40 meV.  For the full SWMC model, due to the breaking of electron-hole
symmetry, the Fermi level can drift within these central Landau subbands, even if the system is at
electroneutrality.  As shown by Yoshioka and Fukuyama \cite{YF81}, due to interaction effects
one then expects a charge density wave (CDW) at sufficiently high fields.  Anomalies in the observed
magnetotransport data corresponding to this CDW transition have indeed been observed
\cite{TAN81}.  The presence of stacking faults, which produce bound states away from the central
Landau levels, should not affect this picture.

However, if the graphite is lightly doped, a different picture emerges \cite{BHRA07}.
In this case, the central Landau levels become filled at a field $B^*=\half n d\,\phi\ns_0$,
where $n$ is the bulk carrier density, $d$ is the interplane separation (\ie\ the $c$-axis
lattice constant is $2d$ due to Bernal stacking).  For fields $B<B^*$, the central Landau
bands are filled, and the Hall conductivity should be quantized at a value
$2e^2/hd$ \cite{BHRA07}.  As $B$ is decreased further, the Fermi level crosses the bound
state energy.  The bound state Landau levels (one for each spin value and inequivalent zone corner)
then makes a contribution to $\sigma\ns_{xy}$, of magnitude $\rmDelta \sigma^\ssr{fault}_{xy}=2x e^2/hd$,
as shown in the sketch in fig. \ref{sketch_QHE}, where $x$ is the concentration of stacking faults.
Upon further reducing $B$, the Fermi level enters
into the first bulk band, and $\sigma\ns_{xy}$ begins to rise continuously.  As $E\ns_\ssr{F}$ crosses
other bound state Landau levels, additional small jumps of $\rmDelta \sigma^\ssr{fault}_{xy}=2x e^2/hd$
should appear.  At a finite concentration $x$ of stacking faults, the bound states will themselves form a band,
and the small jumps will no longer have infinite slope.

The scenario discussed here shows how anomalous features could occur in the high
field magnetotransport of doped graphite, however we cannot find any obvious connection
between our work and the observations of  Kempa \etal\ \cite{KEK06}.

Stacking defects below a graphite surface decouple the surface
region from the bulk, leading to quasi-two-dimensional behavior,
with localized Landau levels.  We have shown how such buried defects leave a signature
which can be measured by surface spectroscopy.

Finally, our results suggest that the electronic properties of few
layer graphene samples can be substantially modified by changes in
the stacking order.

\section{Acknowledgments} The authors gratefully acknowledge conversations with A. Bernevig,
P. Esquinazi, M. Fogler, N. Garc{\'\i}a, T. Hughes, and S. Raghu.
This work was supported by MEC (Spain) through grant
FIS2005-05478-C02-01 and CONSOLIDER CSD2007-00010, the Comunidad de
Madrid, through CITECNOMIK, CM2006-S-0505-ESP-0337, the EU Contract
12881 (NEST).

\section{Appendix : Full SWMC Treatment of Stacking Fault}
We define the vectors
\begin{equation}
\psi\ns_n=\begin{pmatrix} u^\alpha_n \\ v^\alpha_n \\ u^\beta_n \\ v^\beta_n \end{pmatrix} \ (n<0)  \qquad,\qquad
\phi\ns_n=\begin{pmatrix} v^\gamma_n \\ u^\gamma_n \\ v^\beta_n \\ u^\beta_n \end{pmatrix} \ (n>0)\ .
\end{equation}

\begin{figure}[!t]
\centering
\includegraphics[width=7.5cm]{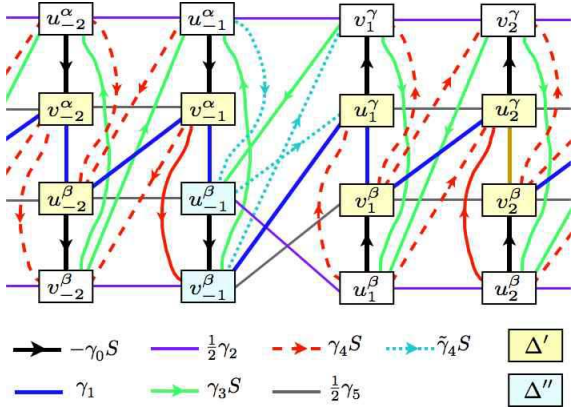}
\caption
{\label{stackfig_b} SWMC couplings for a stacking defect in Bernal graphite,
showing more clearly the four sublattice structure on either side of the defect.}
\end{figure}

For a stacking defect $ABABCBCB$ the SWMC couplings are depicted in fig. \ref{stackfig_b}.  In fact,
additional couplings must be introduced at the defect.  In the bulk, sites have either zero or two
$c$-axis neighbors, but at the stacking fault there are two sites with a single such neighbor.
One expects the associated on-site energy $\rmDelta''\approx\half\rmDelta$.  In addition, there
are three interlayer couplings at the defect which in principle are distinct from $\gth$ and $\gfo$,
and which we denote in the figure by dotted pale blue lines, with hopping amplitude
${\tilde\gamma}\ns_4$.  For simplicity, we shall take $\rmDelta''=\rmDelta$
and ${\tilde\gamma}\ns_4=\gfo$ for two of the links, and ${\tilde\gamma}\ns_4=\gth$ for the
other link.  For details, see the definition of the $F$ matrix below.

 Let each pair of layers be indexed by a nonzero integer $n$.
From the figures, we can read off the Schr{\"o}dinger equations
\begin{align}
M\psi\nd_{n-1} + K\psi\nd_n + M\yd\psi\nd_{n+1}&=0 \quad (n<-1)\\
M^*\phi\nd_{n-1} + K^*\phi\nd_n + M^\rmt\phi\nd_{n+1}&=0  \quad (n>1)\ ,
\end{align}
where, consistent with the full SWMC Hamiltonian \cite{SWMC,DD02},
\begin{equation}
K=\begin{pmatrix} -E & -\gze\,S & \gfo\,S & \gth\,S^* \\ -\gze\,S^* & -E+\dbi & \gon & \gfo\,S \\
\gfo\,S^* & \gon & -E+\dbi & -\gze\,S \\ \gth\,S & \gfo\,S^* & -\gze\,S^* & -E \end{pmatrix}
\end{equation}
and
\begin{equation}
M=\begin{pmatrix} \half\gtw & 0 & \gfo\,S & \gth\,S^* \\
0 & \half\gfi & \gon & \gfo\,S \\ 0 & 0 & \half\gfi & 0 \\ 0 & 0 & 0 & \half\gtw \end{pmatrix}\ .
\end{equation}
We take the SWMC parameters from ref. \cite{DD02}
\begin{align*}
\gze&=3160\,{\rm meV} & \gon &= 390\,{\rm meV}\\
\gtw&=-20\,{\rm meV} & \gth&=315\,{\rm meV}\\
\gfo& =44\,{\rm meV}& \gfi&=38\,{\rm meV}\ ,
\end{align*}
with $\rmDelta=-8\,$meV.
Here, $\dbi$ is a combination of the original SWMC parameters:
\begin{equation}
\dbi=\rmDelta + \gfi-\gtw\ ,
\end{equation}
hence $\dbi=50\,$meV.

At the defect, the Schr{\"o}dinger equation yields
\begin{align}
M\psi\nd_{-2} + K\psi\nd_{-1} + F\yd\phi\nd_1&=0\label{defecta}\\
F\psi\nd_{-1} + K^*\phi\ns_1 + M^\rmt\phi\nd_2&=0\ ,\label{defectb}
\end{align}
where
\begin{equation}
F=\begin{pmatrix} \half\gtw & 0 & \gth\,S &\gfo\,S^* \\
0 & 0 & \gfo \,S^* & \gon \\ 0 & 0 & 0 & \half\gfi \\ 0 & 0 & \half\gtw & 0 \end{pmatrix}\ .
\end{equation}

\subsection{Scattering matrix and bound states}
We write $\psi\nd_n=z^n\,\xhi$ (for $n<0$) and $\phi\nd_n={z^*}^n\,\xhi^*$ (for $n>0$).
In the bulk ($|n|>0$), we then have (for both sides)
\begin{equation}
\big(z^{-1}M + K + z \,M\yd\big)\,\xhi=0\ .
\label{chieqn}
\end{equation}
In order for a solution to exist, we require
\begin{equation}
P(z)\equiv{\rm det}\,\big(z^{-1}M + K + z \,M\yd\big)=0\ ,
\end{equation}
which is an eighth order equation in $z$.  Note that $P(z)=0$ guarantees that $P({z^*}^{-1})=0$.
It can also be shown, due to the form of $M$, that $P(z)=P(z^{-1})$.  Thus, the allowed values of $z$
come in sets $(z,z^*, z^{-1},{z^*}^{-1})$.

Within a bulk energy band, two of the eight $z$ roots are unimodular, and may be written as
$z\ns_1=e^{ i k}$ and $z\ns_5=e^{-ik}$ with $k$ real.  Their associated eigenvectors are
$\xhi\ns_{1,5}$.  Of the remaining six roots, three ($z\ns_2$, $z\ns_3$, $z\ns_4$) each have
modulus greater than unity and are thus unnormalizable on the right.  The remaining three roots
($z\ns_6$, $z\ns_7$, $z\ns_8$) each have modulus smaller than unity and are unnormalizable
on the left.  We keep only the normalizable solutions and write
\begin{align}
n&<0\ :& \psi\ns_n&=\cI\,e^{ikn}\,\xhi\ns_1 + \cO'\,e^{-ikn}\,\xhi\ns_5\\
&&&\qquad+ A\ns_2\,z_2^n\,\xhi\ns_3 +  A^n_3\,z^n_3\,\xhi\ns_3
 +  A\ns_4\,z^n_4\,\xhi\ns_4\nonumber\\
&&&\nonumber\\
n&>0\ :& \phi\ns_n&=\cI'\,e^{-ikn}\,\xhi^*_1 + \cO\,e^{ikn}\,\xhi^*_5 \\
&&&\qquad+  A\ns_6\,{z^*_6}^n\,\xhi^*_6 +  A\ns_7\,{z^*_7}^n\,\xhi^*_7
+ A\ns_8\,{z^*_8}^n\,\xhi^*_8\ .\nonumber
 \end{align}
Equations (\ref{defecta},\ref{defectb}) then yield eight equations in the ten unknowns
$(\cI,\cO,\cI',\cO')$ and $(A\ns_2,A\ns_3,A\ns_4,A\ns_6,A\ns_7,A\ns_8)$.  These then
determine, for each energy $E$, the $\cS$-matrix, defined by the relation
\begin{equation}
\begin{pmatrix} \cO \\ \cO' \end{pmatrix} = \stackrel{\cS}{\overbrace{\begin{pmatrix} t & r' \\ r & t' \end{pmatrix}}}
\begin{pmatrix} \cI \\ \cI' \end{pmatrix}
\end{equation}

If two bands overlap, then we have eigenvalues $z\nd_{1,5}=e^{\pm ik}$ and $z\ns_{2,6}=
e^{\pm i p}$, with $\big|z\ns_{3,4}\big|>1$ and $\big|z\ns_{7,8}\big|<1$.
\begin{align}
n&<0\ :& \psi\ns_n&=\cI\,e^{ikn}\,\xhi\ns_1 + \cO'\,e^{-ikn}\,\xhi\ns_5\\
&&&\quad  +{\tilde\cI}\,e^{ipn}\,\xhi\ns_2 + {\tilde\cO}'\,e^{-ipn}\,\xhi\ns_6\nonumber\\
&&&\qquad+  A^n_3\,z^n_3\,\xhi\ns_3 +A\ns_4\,z^n_4\,\xhi\ns_4\nonumber\\
&&&\nonumber\\
n&>0\ :& \phi\ns_n&=\cI'\,e^{-ikn}\,\xhi^*_1 + \cO\,e^{ikn}\,\xhi^*_5 \\
&&&\quad +{\tilde\cI}'\,e^{-ipn}\,\xhi^*_2+ {\tilde\cO}\,e^{ipn}\,\xhi^*_6\nonumber\\
&&&\qquad +  A\ns_7\,{z^*_7}^n\,\xhi^*_7+ A\ns_8\,{z^*_8}^n\,\xhi^*_8 \ .\nonumber
 \end{align}
The $S$-matrix is then $4\times 4$, and we should take care to properly define it to act on
{\it flux amplitudes\/}, {\it viz.}
\begin{equation}
\begin{pmatrix} v^{1/2}_{1,k}\>\cO \\ v^{1/2}_{1,k}\>\cO' \\
v^{1/2}_{2,p}\>{\tilde\cO} \\ v^{1/2}_{2,p}\>{\tilde\cO}' \end{pmatrix} = \cS
\begin{pmatrix} v^{1/2}_{1,k}\>\cI \\ v^{1/2}_{1,k}\>\cI' \\
v^{1/2}_{2,p}\>{\tilde\cI} \\ v^{1/2}_{2,p}\>{\tilde\cI}' \end{pmatrix} \ ,
\end{equation}
where $v\ns_{1,k}=\pz E\ns_1(k)/\pz k$ and $v\ns_{2,p}=\pz E\ns_2(p)/\pz p$.
If three bands overlap, the $S$-matrix is $6\times 6$.

\subsection{Bound states}
When $E$ does not lie within a bulk band, we write
\begin{align}
n&<0\ :& \psi\ns_n&=A\ns_1\,z^n_1\,\xhi\ns_1 +  A\ns_2\,z^n_2\,\xhi\ns_2\\
&&&\qquad\qquad+ A^n_3\,z\ns_3\,\xhi\ns_3 +A\ns_4 \,z^n_4\,\xhi\ns_4 \nonumber\\
n&>0\ :& \phi\ns_n&= A\ns_5\,{z^*_5}^n\,\xhi^*_5 +  A\ns_6\,{z^*_6}^n\,\xhi^*_6\\
&&&\qquad\qquad+ A\ns_7\,{z^*_7}^n\,\xhi^*_7+A\ns_8\,{z^*_8}^n\,\xhi^*_8\ .\nonumber
\end{align}
Here, $\big|z\ns_{1,2,3,4}\big|>1$ and  $\big|z\ns_{5,6,7,8}\big|<1$.  Without loss of generality, we
may assume
\begin{equation}
z^*_u=z^{-1}_{u+4}\ ,
\end{equation}
for $u=1,2,3,4$.  Equations (\ref{defecta},\ref{defectb}) now give eight
homogeneous equations in the eight unknowns $A\ns_{1-8}$.  A nontrivial solution can only exist when the corresponding
determinant vanishes, which puts a single complex condition on the energy $E$.  The solutions are the allowed bound states.

We now apply eqns. \ref{defecta} and \ref{defectb}:
\begin{align}
M\psi\nd_{-2} + K\psi\nd_{-1} + F\yd\phi\nd_1&=0\\
F\psi\nd_{-1} + K^*\phi\ns_1 + M^\rmt\phi\nd_2&=0
\end{align}
to
\begin{equation}
\psi\nd_n=\sum_{u=1}^4 A\ns_u\,z^n_u\,\xhi\ns_u\quad,\quad
\phi\nd_n=\sum_{l=5}^8 A\ns_l\,{z^*_l}^n\,\xhi^*_l
\end{equation}
using
\begin{align}
\big(z^{-1}M + K + z \,M\yd\big)\,\xhi&=0\\
\big({z^*}^{-1}M^* + K^* + z^* \,M^\rmt\big)\,\xhi^*&=0\ .
\end{align}
This yields
\begin{equation}
M\yd\psi\ns_0 = F\yd\phi\ns_1 \qquad,\qquad F\,\psi\nd_{-1}=M^*\phi\ns_0\ ,
\end{equation}
which, expanded, gives
\begin{equation}
\sum_{u=1}^4 A\ns_u\,M\yd\xhi\ns_u = \sum_{l=5}^8 A\ns_l\,z^*_l\,F\yd \xhi^*_l
\label{BSA}
\end{equation}
and
\begin{equation}
\sum_{u=1}^4 A\ns_u\,z_u^{-1}\,F\xhi\ns_u = \sum_{l=5}^8 A\ns_l\,M^* \xhi^*_l\ .
\label{BSB}
\end{equation}
These give $8$ homogeneous equations in the $8$ unknowns can only be solved when
the corresponding determinant vanishes, which is the condition that $E$ lie at a bound state energy.

\subsection{Method of solution}
Eqn. \ref{chieqn} can be written as two coupled equations,
\begin{align}
z^{-1}\,M\,\xhi + \xhi'&=0\\
K\,\xhi + z\,M\yd\,\xhi -\xhi'&=0\ .
\end{align}
These equations may be recast as the rank-8 system,
\begin{equation}
\begin{pmatrix} z +NK &  -N \\ M & z  \end{pmatrix}
\begin{pmatrix} \xhi \\ \xhi' \end{pmatrix}=0\ ,
\end{equation}
where $N\equiv  {M\yd}^{-1}$.
Thus the solutions $z\ns_j$ are the complex eigenvalues of the matrix
\begin{equation}
Q=\begin{pmatrix} -NK &  N\\ -M & 0  \end{pmatrix}
\begin{pmatrix} \xhi \\ \xhi' \end{pmatrix}\ .
\end{equation}
Note that ${\rm det}(Q)={\rm det}(M)\cdot{\rm det}(N)=1$ independent of $K$
and of the above-diagonal elements of $M$ and the below-diagonal elements of $N$.
From row reduction, it is easy to derive
\begin{equation}
N={4\over\gtw\gfi}\begin{pmatrix} \half\gfi & 0 & 0 & 0 \\
0 &\half\gtw & 0 & 0 \\
-\gfo\,S^* & -\gon\gtw\gamma_5^{-1} & \half\gtw & 0 \\
-\gamma_2^{-1}\gth\gfi\,S &\gfo\,S^* & 0 & \half\gfi\end{pmatrix}\ .
\end{equation}
The equations (\ref{BSA}) and (\ref{BSB}) can now be written as an $8\times 8$ system,
\begin{equation}
\stackrel{R}{\overbrace{
\begin{pmatrix} M\yd_{ab}\,\xi\ns_{bu} & - z^*_l \, F\yd_{ab}\,\xi^*_{bl} \\
&\\
z^{-1}_u \, F\ns_{ab}\,\xi\ns_{bu}& -  M^*_{ab}\,\xi^*_{bl} \end{pmatrix} }}
\begin{pmatrix} A\ns_{u=1,2,3,4} \\  \\ A\ns_{l=5,6,7,8} \end{pmatrix}=0\ ,
\end{equation}
where $a$, $b$, and $u$ run from $1$ to $4$, and $l$ runs from $5$ to $8$.
The implied sums for each submatrix are over $b$ and not $u$ or $l$, and
$\xi\ns_{ij}$ is the matrix of eigenvectors of $Q$:
\begin{equation}
\sum_{k=1}^8 Q\ns_{ik}\,\xi\ns_{kj}=z\ns_i\,\xi\ns_{ij}\quad\hbox{\rm (no sum on $i$)}\ ,
\end{equation}
where $i$, $j$, and $k$ run from $1$ to $8$.  The bound state condition is
${\rm det}(R)=0$.

\begin{figure}[!b]
\centering
\includegraphics[width=8.0cm]{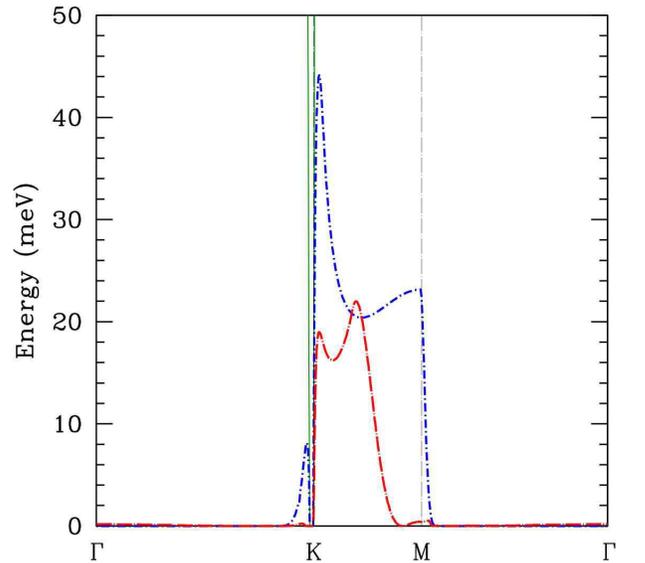}
\caption
{\label{boundstate} Binding energies for bound states within the gap at negative energies (blue short dash - dot curve)
and positive energies (red long dash - dot curve), for the SWMC model with a single stacking fault, as a function of
wavevector in the basal Brillouin zone.   The solid green line shows the energy gap between bonding and antibonding $\pi$ bands,
which collapses in the vicinity of the basal zone corner $K$.}
\end{figure}

We have numerically found the bound states lying in the gap between
the bonding and antibonding $pi$ bands of graphite.  Our results are
displayed in fig. \ref{boundstate}.  For the full SWMC calculation,
there is no longer particle-hole symmetry. We find that the binding
energy (\ie\ the distance of the bound state from the closest band
extremum) is considerable along the entire $KM$ edge.  This is in
contrast to our analytic results for the nearest neighbor model,
where the bound state energy was considerable only for $|S|\approx
\gon/2\gze\simeq 0.062$, which is satisfied only on a small ring
about the $K$ and $K'$ points. On the other hand, the lack of bound
states along the $\Gamma K$ edge implies a finite broadening of the
Landau levels derived from this band.

It is important to realize that the SWMC model itself is only valid close to the $K$-$H$ spine in the Brillouin zone.  The model must
be extended, as in ref. \cite{JD73}, to include other tight binding parameters, in order to fit the $\pi$ band throughout the entire zone,
which is necessary in order to model various optical transitions.  In this case, the in-plane hopping is modified:
\begin{equation}
\gze S \to \gamma^\sss{(1)}_0\,S\ns_1+\gamma^\sss{(2)}_0\,S\ns_2+\gamma^\sss{(3)}_0\,S\ns_3\ ,
\end{equation}
where $\gamma^\sss{(n)}$ and $S\ns_n$ are, respectively, the amplitude and lattice sum of $e^{i\bfk\cdot\bfdelta}$
corresponding to the $n^\ssr{th}$ nearest neighbor in-plane inter-sublattice hopping  \cite{JD73}, subject to the constraint
\begin{equation}
\gamma_0^\ssr{SWMC}=\gamma^\sss{(1)}_0 - 2\,\gamma^\sss{(2)}_0 + \gamma^\sss{(3)}_0\ .
\end{equation}
It is a rather simple matter to include such effects in our calculation, and we find in general, for a broad set of possible
parameterizations satisfying the constraint, that our results have the same qualitative features.

Our approximations regarding the parameters $\rmDelta''$ and ${\tilde\gamma}\ns_4$ are such that, were
their values known, our binding energies could easily be off by perhaps a few tens of millivolts.
We expect, however, that the general features found here should still pertain, namely a single bound
state whose binding energy is maximized at several tens of millivolts along the $K$-$M$ edge in the basal
Brillouin zone.

\end{document}